\documentclass[traditabstract,longauth]{aa}  
\usepackage{graphicx}
\usepackage{tabularx}
\usepackage{multirow}
\usepackage[colorlinks=true, linkcolor=blue, citecolor=blue, draft=False]{hyperref}
\usepackage{gensymb}
\usepackage{color}
\usepackage{txfonts}
\usepackage{amsmath}
\usepackage{ulem}

\defcitealias{Franco2018}{F18}
\defcitealias{Franco2020}{F20b}

\begin{document} 
\title{GOODS-ALMA: Using IRAC and VLA to probe fainter millimeter galaxies}
\author{M.~Franco\inst{\ref{inst1},\ref{inst2}}\thanks{Email: m.franco@herts.ac.uk}
\and D.~Elbaz\inst{\ref{inst1}} 
\and L.~Zhou\inst{\ref{inst1},\ref{inst3},\ref{inst4}}
\and B.~Magnelli\inst{\ref{inst5}}
\and C.~Schreiber\inst{\ref{inst6}}
\and L.~Ciesla\inst{\ref{inst1},\ref{inst7}}
\and M.~Dickinson\inst{\ref{inst8}}
\and N.~Nagar\inst{\ref{inst9},\ref{inst10}}
\and G.~Magdis\inst{\ref{inst11},\ref{inst12},\ref{inst13},\ref{inst14}}
\and D.~M.~Alexander\inst{\ref{inst15}}
\and M.~B\'ethermin\inst{\ref{inst7}}
\and R.~Demarco\inst{\ref{inst9},\ref{inst10}}
\and E.~Daddi\inst{\ref{inst1}}
\and T.~Wang\inst{\ref{inst1},\ref{inst16}}
\and J.~Mullaney\inst{\ref{inst17}}
\and H.~Inami\inst{\ref{inst18},\ref{inst19}}
\and X.~Shu\inst{\ref{inst20}}
\and F.~Bournaud\inst{\ref{inst1}}
\and R.~Chary\inst{\ref{inst21}}
\and R.~T.~Coogan\inst{\ref{inst22}}
\and H.~Ferguson\inst{\ref{inst23}}
\and S.~L.~Finkelstein\inst{\ref{inst24}}
\and M.~Giavalisco\inst{\ref{inst25}}
\and C.~G\'omez-Guijarro\inst{\ref{inst1}}
\and D.~Iono\inst{\ref{inst26},\ref{inst27}}
\and S.~Juneau\inst{\ref{inst1},\ref{inst8}}
\and G.~Lagache\inst{\ref{inst7}}
\and L.~Lin\inst{\ref{inst28}}
\and K.~Motohara\inst{\ref{inst29}}
\and K.~Okumura\inst{\ref{inst1}}
\and M.~Pannella\inst{\ref{inst30},\ref{inst31}}
\and C.~Papovich\inst{\ref{inst32},\ref{inst33}}
\and A.~Pope\inst{\ref{inst25}}
\and W.~Rujopakarn\inst{\ref{inst34},\ref{inst35},\ref{inst36}}
\and J.~Silverman\inst{\ref{inst36}}
\and M.~Xiao\inst{\ref{inst1},\ref{inst3}}
}

\institute{AIM, CEA, CNRS, Universit\'{e} Paris-Saclay, Universit\'{e} Paris Diderot, Sorbonne Paris Cit\'{e}, F-91191 Gif-sur-Yvette, France\label{inst1}
\and Centre for Astrophysics Research, University of Hertfordshire, Hatfield, AL10 9AB, UK\label{inst2}
\and School of Astronomy and Space Science, Nanjing University, Nanjing 210093, China\label{inst3}
\and Key Laboratory of Modern Astronomy and Astrophysics (Nanjing University), Ministry of Education, Nanjing 210093, China\label{inst4}
\and Argelander-Institut f\"{u}r Astronomie, Universit\"{a}t Bonn, Auf dem H\"{u}gel 71, D-53121 Bonn, Germany\label{inst5}
\and Department of Physics, University of Oxford, Keble Road, Oxford OX1 3RH, UK\label{inst6}
\and Aix Marseille Universit\'{e}, CNRS, LAM, Laboratoire d'Astrophysique de Marseille, Marseille, France\label{inst7}
\and Community Science and Data Center/NSF's NOIRLab, 950 N. Cherry Ave., Tucson, AZ 85719, USA \label{inst8}
\and Department of Astronomy, Universidad de Concepci\'{o}n, Casilla 160-C Concepci\'{o}n, Chile \label{inst9}
\and Departamento de Astronom\'ia, Facultad de Ciencias F\'isicas y Matem\'aticas,
Universidad de Concepci\'on, Concepci\'on, Chile \label{inst10}
\and Cosmic Dawn Center at the Niels Bohr Institute, University of Copenhagen and DTU-Space, Technical University of Denmark\label{inst11}
\and DTU Space, National Space Institute, Technical University of Denmark, Elektrovej 327, DK-2800 Kgs. Lyngby, Denmark\label{inst12}
\and Niels Bohr Institute, University of Copenhagen, DK-2100 Copenhagen, Denmark\label{inst13}
\and Institute for Astronomy, Astrophysics, Space Applications and Remote Sensing, National Observatory of Athens, 15236, Athens, Greece\label{inst14}
\and Centre for Extragalactic Astronomy, Department of Physics, Durham University, Durham DH1 3LE, UK\label{inst15}
\and Institute of Astronomy, University of Tokyo, 2-21-1 Osawa, Mitaka, Tokyo 181-0015, Japan\label{inst16}
\and Department of Physics and Astronomy, The University of Sheffield, Hounsfield Road, Sheffield S3 7RH, UK\label{inst17}
\and Univ Lyon, Univ Lyon1, ENS de Lyon, CNRS, Centre de Recherche Astrophysique de Lyon (CRAL) UMR5574, F-69230, Saint-Genis-Laval, France\label{inst18}
\and Hiroshima Astrophysical Science Center, Hiroshima University, 1-3-1 Kagamiyama, Higashi-Hiroshima, Hiroshima 739-8526, Japan\label{inst19}
\and Department of Physics, Anhui Normal University, Wuhu, Anhui, 241000, China\label{inst20}
\and Infrared Processing and Analysis Center, MS314-6, California Institute of Technology, Pasadena, CA 91125, USA\label{inst21}
\and Max-Planck-Institut f\"{u}r Extraterrestrische Physik (MPE), Giessenbachstr.1, 85748 Garching, Germany \label{inst22}
\and Space Telescope Science Institute, 3700 San Martin Drive, Baltimore, MD 21218, USA\label{inst23}
\and Department of Astronomy, The University of Texas at Austin, Austin, TX 78712, USA\label{inst24}
\and Astronomy Department, University of Massachusetts, Amherst, MA 01003, USA\label{inst25}
\and National Astronomical Observatory of Japan, National Institutes of Natural Sciences, 2-21-1 Osawa, Mitaka, Tokyo 181-8588, Japan\label{inst26}
\and SOKENDAI (The Graduate University for Advanced Studies), 2-21-1 Osawa, Mitaka, Tokyo 181-8588, Japan\label{inst27}
\and Institute of Astronomy \& Astrophysics, Academia Sinica, Taipei 10617, Taiwan\label{inst28}
\and Institute of Astronomy, Graduate School of Science, The University of Tokyo, 2-21-1 Osawa, Mitaka, Tokyo 181-0015, Japan\label{inst29}
\and Astronomy Unit, Department of Physics, University of Trieste, via Tiepolo 11, I-34131 Trieste, Italy\label{inst30}
\and Fakult\"{a}t f\"{u}r Physik der Ludwig-Maximilians-Universit\"{a}t, D-81679 M\"{u}nchen, Germany\label{inst31}
\and Department of Physics and Astronomy, Texas A\&M University, College Station, TX, 77843-4242, USA\label{inst32}
\and George P. and Cynthia Woods Mitchell Institute for Fundamental Physics and Astronomy, Texas A\&M University, College Station, TX, 77843-4242, USA\label{inst33}
\newpage
\and Department of Physics, Faculty of Science, Chulalongkorn University, 254 Phayathai Road, Pathumwan, Bangkok 10330, Thailand\label{inst34}
\and National Astronomical Research Institute of Thailand (Public Organization), Donkaew, Maerim, Chiangmai 50180, Thailand\label{inst35}
\and Kavli Institute for the Physics and Mathematics of the Universe (WPI), The University of Tokyo Institutes for Advanced Study, The University of Tokyo, Kashiwa, Chiba 277-8583, Japan \label{inst36}
}
\date{Received: 30 April 2020; accepted: 07 August 2020}

\abstract{In this paper, we extend the source detection in the GOODS-ALMA field (69 arcmin$^2$, 1$\sigma$\,$\simeq$\,0.18\,mJy\,beam$^{-1}$) to deeper levels than presented in our previous work. Using positional information at 3.6 and 4.5\,$\mu$m (from \textit{Spitzer}-IRAC) as well as the Very Large Array (VLA) at 3GHz, we explore the presence of galaxies detected at 1.1\,mm with ALMA below our original blind detection limit of 4.8-$\sigma$, at which the number of spurious sources starts to dominate over that of real sources. In order to ensure the most reliable counterpart association possible, we have investigated the astrometry differences between different instruments in the GOODS--South field. In addition to a global offset between the Atacama Large Millimeter/submillimeter Array (ALMA) and the \textit{Hubble} Space Telescope (HST) already discussed in previous studies, we have highlighted a local offset between ALMA and the HST that was artificially introduced in the process of building the mosaic of the GOODS--South image. We created a distortion map that can be used to correct for these astrometric issues. In this Supplementary Catalog, we find a total of 16 galaxies, including two galaxies with no counterpart in HST images (also known as optically dark galaxies), down to a 5$\sigma$ limiting depth of H\,=\,28.2 AB (HST/WFC3 F160W). This brings the total sample of GOODS-ALMA 1.1\,mm sources to 35 galaxies. Galaxies in the new sample cover a wider dynamic range in redshift ($z$\,=\,0.65 -- 4.73), are on average twice as large (1.3 vs 0.65 kpc), and have lower stellar masses (M$_{\star}^{\rm SC}$\,=\,7.6$\times$10$^{10}$M$_\odot$ vs M$_{\star}^{\rm MC}$\,=\,1.2$\times$10$^{11}$M$_\odot$). Although exhibiting larger physical sizes, these galaxies still have far-infrared sizes that are significantly more compact than inferred from their optical emission. }


    \maketitle
%

\section{Introduction}

The formation and evolution of the most massive galaxies (M$_{\star}$\,$>$\,5$\times$10$^{10}$\,M$_{\odot}$) at redshifts $z$\,$>$\,2 is still largely debated. Their observed number density exceed theoretical expectations assuming typical dark matter to stellar mass ratios \citep{Steinhardt2016}. The downsizing of galaxy formation challenges theoretical models that match either the low or high mass end but are unable to match both ends \cite[e.g.,][]{Fontanot2009}. The presence of a population of massive passive galaxies at $z$\,$\sim$2 with compact stellar surface densities challenges searches for their progenitors \citep{Van_der_wel2014}. 

As infrared (IR) wavelengths contribute to approximately half of the total extragalactic background light (EBL; e.g., \citealt{Dole2006}), the study of dust-enshrouded star formation in distant galaxies is an important tool to advance our understanding of the evolution of massive galaxies. The first submillimeter extragalactic surveys \citep{Smail1997, Barger1998, Hughes1998} performed with the Submillimetre Common-User Bolometer Array (SCUBA; \citealt{Holland1999}) on the James Clerk Maxwell Telescope (JCMT) have revealed a population of high-redshift galaxies that are massive, highly obscured, and have high star-formation rates (SFRs; see \citealt{Casey2014_review} for a review). Recent observations using the Atacama Large Millimeter/submillimeter Array (ALMA), which provides a spatial resolution more than an order of magnitude higher than SCUBA, have since refined our understanding of galaxy evolution by securing the identification of optical counterparts and allowing us to detect not only extreme galaxies (galaxies with particularly high SFRs, e.g., starburst or lensed galaxies), but also ``normal'' galaxies that are secularly forming stars. It also allows us to resolve out a number of single-dish sources into multiple components  (see \citealt{Hodge2013}).

This paper extends our previous analysis (\citealt{Franco2018}, hereafter \citetalias{Franco2018}) of a deep continuum 1.1\,mm survey with ALMA over an area of 69 arcmin$^2$. This survey is located in the Great Observatories Origins Deep Survey--South (GOODS--South) at a location covered with the deepest integrations in the $H$-band with the HST-WFC3 camera. In \citetalias{Franco2018}, we limited our analysis to the blind detection of ALMA sources without considering other wavelengths. Due to the large number of independent beams in the high-resolution ALMA image, we were limited to sources with a S/N greater than 4.8. Here we extend the detection limit to 3.5-$\sigma$ by cross-matching the ALMA detections with catalogs in the near and mid-IR. The need for a good astrometric calibration led us to introduce an improved correction for the astrometry of the HST image of the GOODS--South field.

This paper is organized as follows: In $\S$\ref{Sect::Data}, we present the data used. 
 In $\S$\ref{Sect::Astrometric_correction}, we describe the astrometric correction to be applied to HST positions to align them with those of ALMA. We give the astrometric correction to be applied for all galaxies in the GOODS--South field present in the \cite{Guo2013} catalog, provided as an external link.
In $\S$\ref{Sect::Catalog_creation}, we present the criteria and methods used to select the sample of galaxies that constitutes the Supplementary Catalog to the Main Catalog presented in \citetalias{Franco2018}.
 In $\S$\ref{Sect::Catalog}, we present the properties of the galaxies of the Supplementary Catalog, including two optically dark galaxies. 
 Finally, in $\S$\ref{Sect::Properties}, we perform a comparative analysis of the distribution of stellar masses, redshifts, and sizes between the sample of galaxies presented in this paper and in \citetalias{Franco2018}  and discuss the implications on the nature of the ALMA sources.
 
Throughout this paper, we adopt a spatially flat $\Lambda$CDM cosmological model with H$_0$\,=\,70 km\,s$^{-1}$Mpc$^{-1}$, $\Omega_m$\,=\,0.3 and $\Omega_{\Lambda}$\,=\,0.7. We assume a Salpeter \citep{Salpeter1955} initial mass function (IMF). We use the conversion factor of M$_\star$ (\citealt{Salpeter1955} IMF)\,=\,1.7\,$\times$\,M$_\star$ (\citealt{Chabrier2003} IMF). All magnitudes are quoted in the AB system \citep{Oke1983}.

\section{Data}\label{Sect::Data}

\subsection{ALMA data}\label{Sect::ALMAobs}
This paper uses the 1.1\,mm photometric survey obtained with the Atacama Large Millimeter/Submillimeter Array (ALMA) between August and September 2016 (Project ID: 2015.1.00543.S; PI: D. Elbaz). The survey performed using band 6 covers an effective area of 69 arcmin$^2$ matching the deepest HST/WFC3 $H$-band observation taken as part of the Cosmic Assembly Near-infrared Deep Extragalactic Legacy Survey (CANDELS; \citealt{Grogin2011}, \citealt{Koekemoer2011}, PIs: S. Faber, H. Ferguson), in the  GOODS--South field. It is centered at $\alpha$\,=\,3$^{\rm h}$ 32$^{\rm m}$ 30.0$^{\rm s} $, $\delta$\,=\,-27$\degree$ 48$\arcmin$ 00$\arcsec$ (J2000). The original 0\farcs2 angular resolution was tapered with a homogeneous and circular synthesized beam of $0\farcs60$ Full-Width Half Maximum (FWHM; hereafter 0\farcs60-mosaic). 
The sensitivity of the 0\farcs60-mosaic varying only slightly within the six slices of the survey around a median value of $\simeq$ 0.18\,mJy beam$^{-1}$, we consider to first order to work with homogeneous coverage and we do not distinguish, unless otherwise stated, the specifics of each slice.

\subsection{Additional data}\label{Sect::add_data}

\subsubsection{IRAC catalog}\label{Sect::IRAC_data}
We use the \textit{Spitzer}-Cosmic Assembly Near-Infrared Deep Extragalactic Survey (\textit{S}--CANDELS; \citealt{Ashby2015}) catalog of galaxies detected at 3.6 and 4.5\,$\mu$m with the Infrared Array Camera (\text{IRAC}; \citealt{Fazio2004}) aboard the \textit{Spitzer} Space Telescope \citep{Werner2004}. The catalog \citep{Ashby2015} -- hereafter S-CANDELS catalog -- that reaches a 5-$\sigma$ depth of 26.5 mag (AB) includes  2627 galaxies in the GOODS-ALMA field (i.e., approximately 38 sources/arcmin$^{2}$.

\subsubsection{Near-infrared $K_s$-band catalog}\label{Sect::NIR_data}

We use the 2.2\,$\mu$m catalog described in \cite{Straatman2016} that uses an ultradeep image resulting from the combination of multiple observations in the $K$ and $K_s$ bands from: 
\textit{(i)} the Very Large Telescope (VLT), which combines the images of GOODS--South done with the Infrared Spectrometer and array camera (ISAAC; \citealt{Moorwood1999}) in the $K_s$-band \citep{Retzlaff2010} with the High Acuity Wide-field $K$-band imager (Hawk-I; \citealt{Kissler2008}) image in the $K$-band \citep{Fontana2014}, 
\textit{(ii)} the 6.5m Magellan Baade Telescope combining the $K_s$-band image from the \texttt{FourStar} Galaxy Evolution Survey (ZFOURGE, PI: I. Labb\'{e}) using the \texttt{FourStar} near-infrared Camera \citep{Persson2013} with the $K$-band image using the Persson's Auxillary Nasmyth infrared camera (PANIC; \citealt{Martini2004}) in the HUDF (PI: I. Labb\'e),
\textit{(iii)} the Canada-France-Hawaii Telescope (CFHT), with the $K$-band image done with the wide-field infrared camera (WIRCam; \citealt{Puget2004}) as part of the Taiwan ECDFS Near Infrared Survey (TENIS; \citealt{Hsieh2012}). The 5-$\sigma$ point-source detection threshold in this ultradeep $K_s$ image reaches a magnitude between 26.2 and 26.5, which leads to an average galaxy surface density of approximately 168 sources/arcmin$^{2}$.

\subsubsection{Radio catalog}\label{Sect::radio_data}

A radio image that encompasses the GOODS-ALMA field was observed with the Karl G. Jansky Very Large Array (VLA) at a frequency of 3 GHz (10\,cm) and an angular resolution of $\sim$0$\farcs$3 for a total of 177 hours (configurations A, B, \& C; PI: W. Rujopakarn). Down to the average depth of the radio catalog within the GOODS-ALMA region of RMS\,=\,2.1\,$\mu$Jy.beam$^{-1}$, the average surface density of radio sources is approximately 5 sources/arcmin$^{2}$.

\subsubsection{HST $H$-band catalog}\label{Sect::HST_data}

The GOODS-ALMA area covers the deepest $H$-band part of the Cosmic Assembly Near-IR Deep Extragalactic Legacy Survey (CANDELS; \citealt{Grogin2011}) field (central one-third of the field). The point source catalog reaches a 5-$\sigma$ depth of 28.16 mag (AB) in the $H_{160}$ filter (measured within a fixed aperture of 0\farcs17; \citealt{Guo2013}). The surface density of galaxies detected at 1.6\,$\mu$m with the Wide Field Camera 3 / infrared channel (WFC3/IR) within the GOODS-ALMA field is approximately 233 sources/arcmin$^{2}$. We also cross-checked missing sources against the catalogs of \cite{Koekemoer2011} and \cite{Skelton2014}.

\begin{figure*}
\centering
\begin{minipage}[t]{1.\textwidth}
\resizebox{\hsize}{!} {
\includegraphics[width=1.3cm,clip]{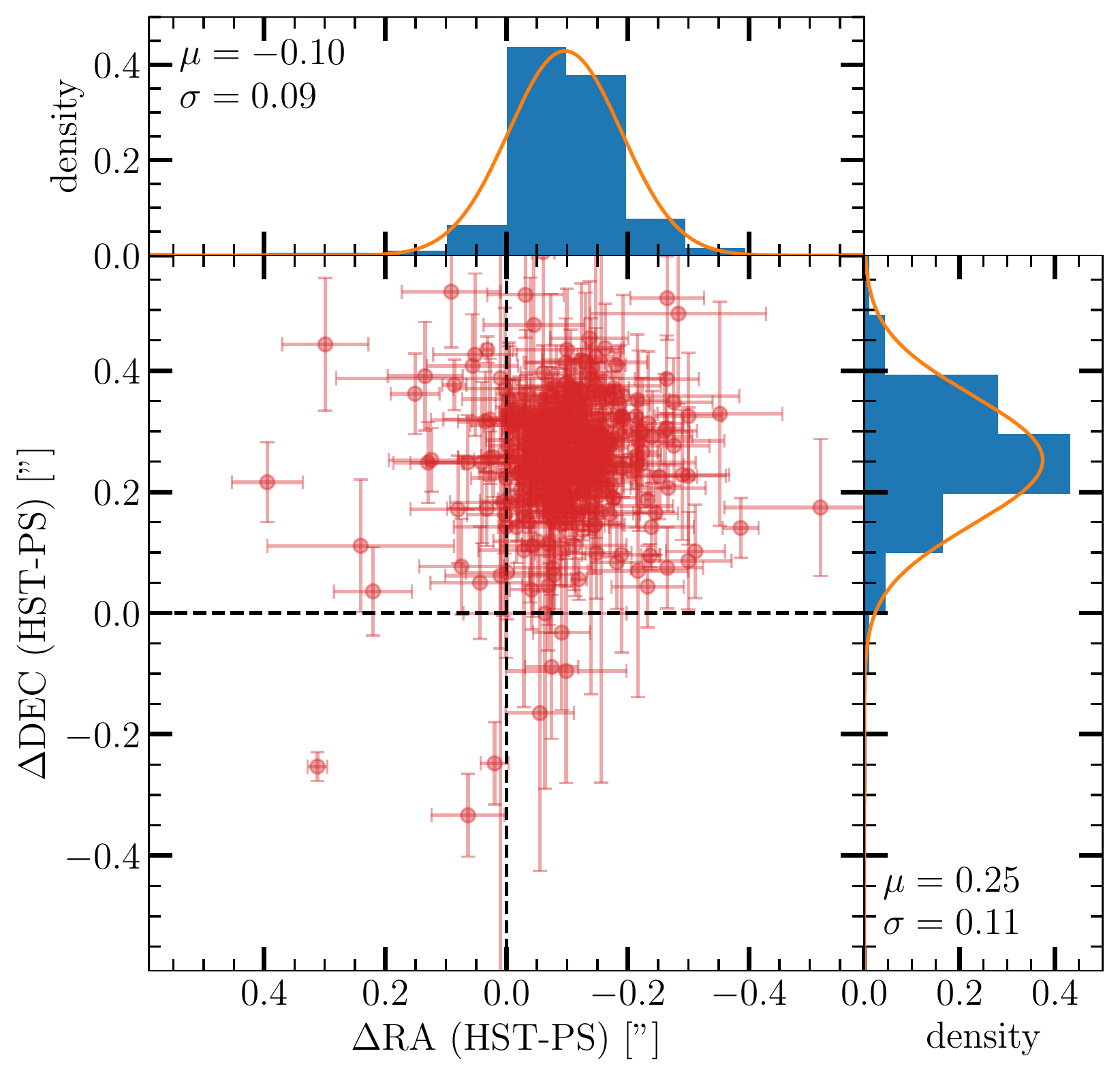}\\
\includegraphics[width=1.3cm,clip]{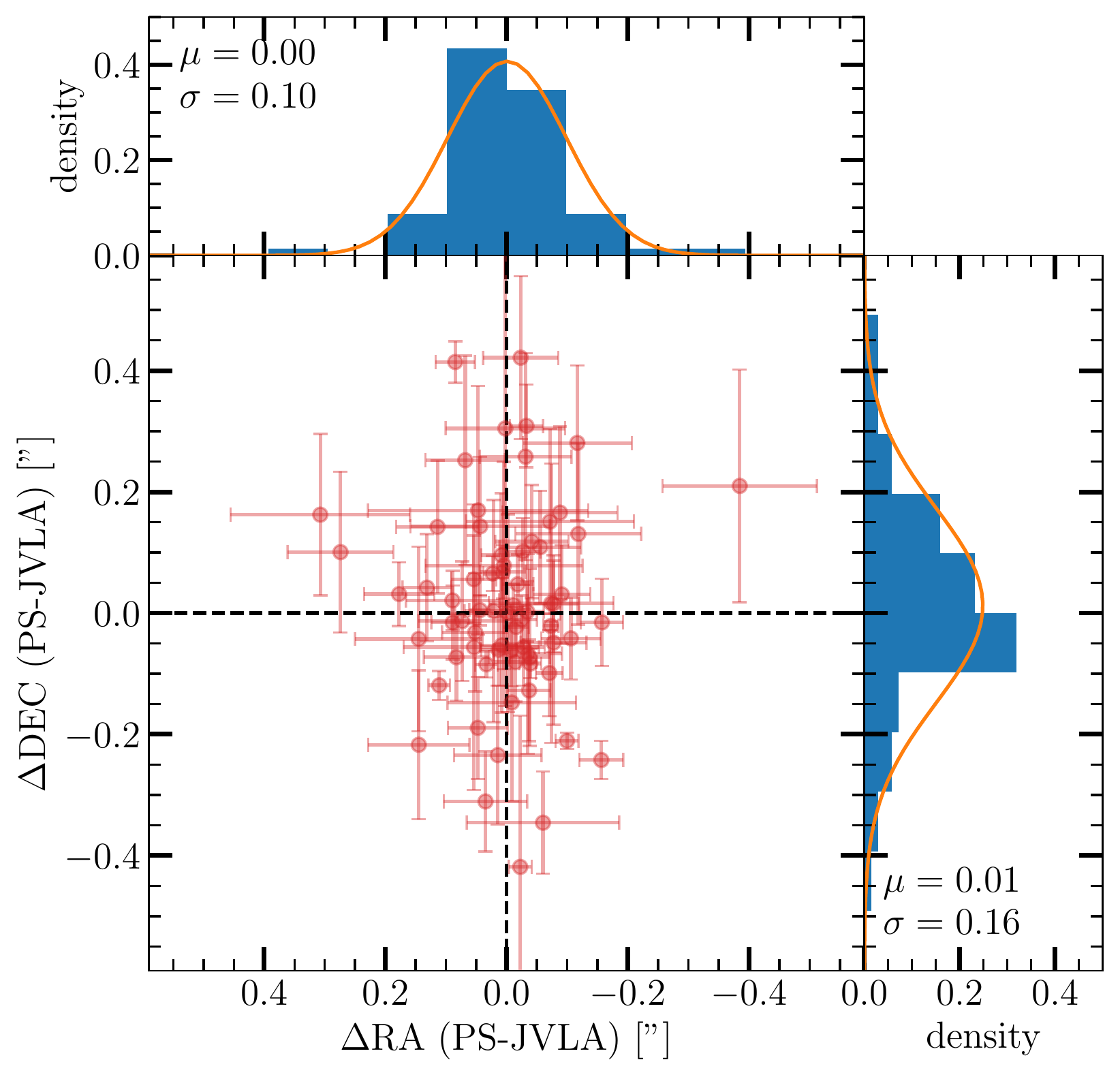}\\
\includegraphics[width=1.3cm,clip]{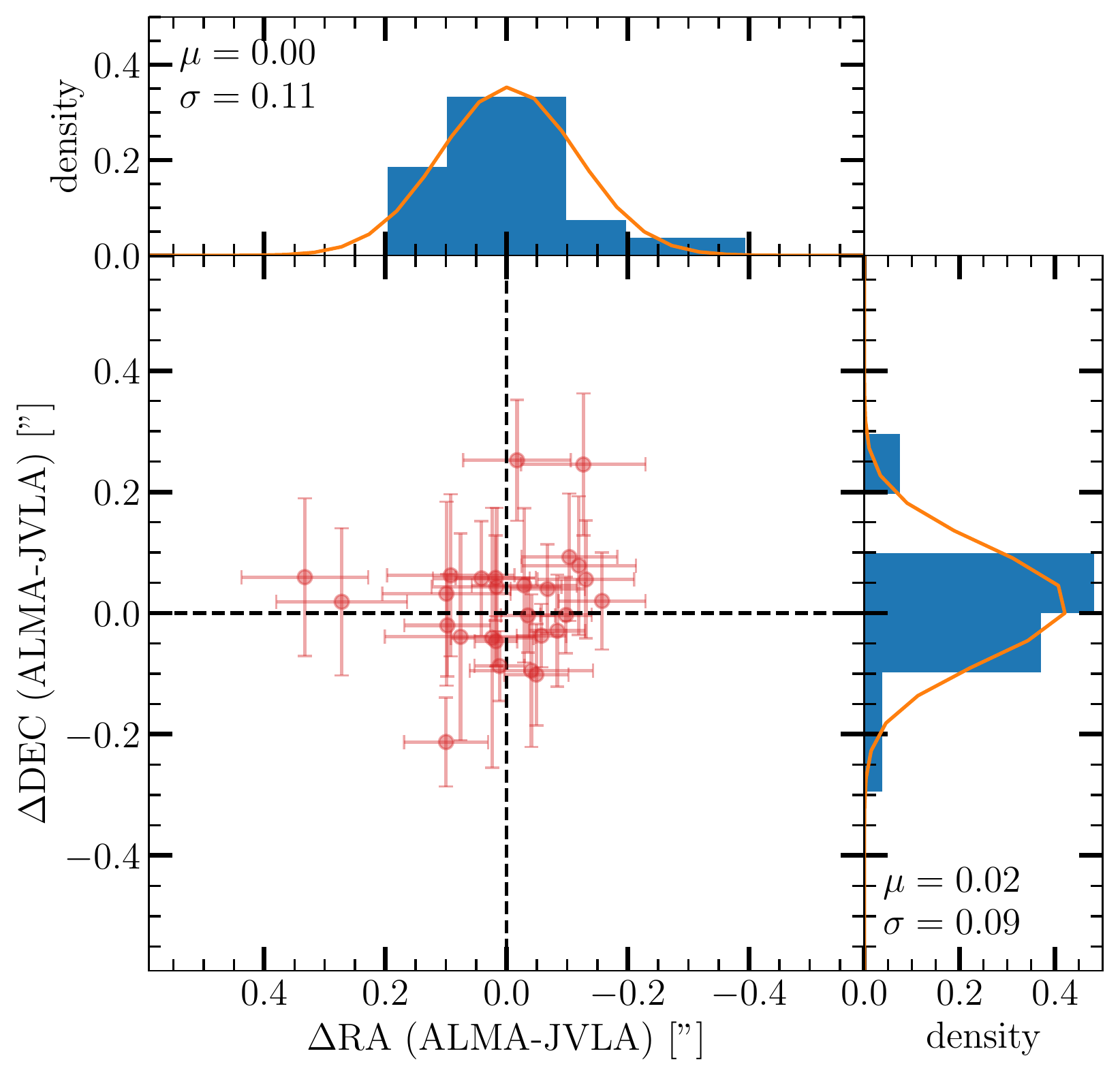}\\
}
\end{minipage}
\caption{Left: offset between Pan-STARRS (PS) and the HST;
Middle: offset between the VLA and Pan-STARRS;
Right: offset between the VLA and ALMA.
For each panel, the histogram of the offsets in RA and Dec is shown as well as a fit with a Gaussian function (orange curve). The position of the peak and the standard deviation of the Gaussian is indicated for each curve.
The middle and the right panels show that there are no significant astrometric differences between ALMA and the VLA nor between the VLA and Pan-STARRS, while the left panel shows a clear shift in both RA and Dec between the positions of 375 sources in common between the Pan-STARRS and HST images. We measure a systematic offset of $\Delta$RA\,=\,$-$96\,$\pm$\,83 mas and $\Delta$Dec\,=\,252\,$\pm$\,107 mas. In addition, a local offset is presented in Fig.~\ref{Systematic_offset}. }
\label{offset}
\end{figure*}

   \begin{figure*}
   \centering
   \includegraphics[width=0.7\hsize]{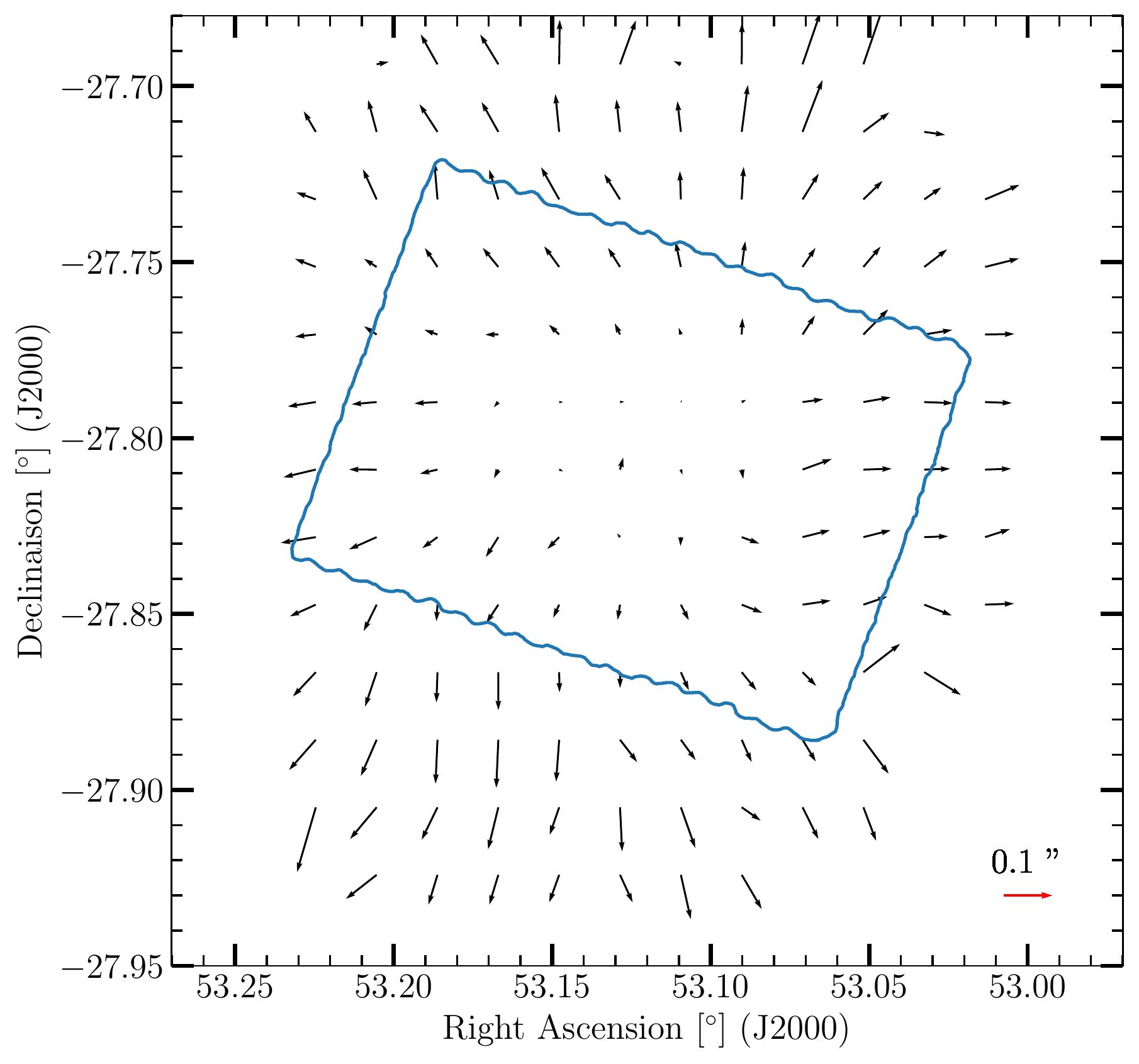}
\caption{Difference between the HST and Pan-STARRS position after subtraction of the median offset value. Each arrow represents a sliding median including on average 15 galaxies, with an overlap of 60 percent between neighboring arrows. The blue line defines the area encompassing the GOODS-ALMA survey.}
         \label{Systematic_offset}
   \end{figure*}

\section{Astrometric correction of the HST image of GOODS--South}\label{Sect::Astrometric_correction}
We describe in \citetalias{Franco2018}  the presence of a systematic offset of $\Delta$RA\,=\,$-$96\,$\pm$\,113 mas in right ascension, and $\Delta$Dec\,=\,261\,$\pm$\,25 mas in declination between the ALMA and HST images. This offset, interpreted as a positional shift of the HST image with respect to all other reference frames, is in good agreement with the offset previously discussed in \cite{Dunlop2017} and \cite{Rujopakarn2016}. However, this correction has not been made to the HST v2.0 release for GOODS--South, in part because no external astrometric reference data with both suitable absolute accuracy and faint source density (such as the SDSS) were available\footnote{\url{https://archive.stsci.edu/pub/hlsp/goods/v2/h\_goods\_v2.0\_rdm.html}}.

The offset used until now only corrects for the bulk global shift in astrometry but it does not account for the relative error in the astrometric calibration that was introduced in the building of the HST mosaic. In the following, we propose to determine this local correction that behaves like a distortion correction. This correction is important in the present study since we aim at using knowledge of existing sources from other wavelengths in order to push the ALMA detection limit to deeper levels.

We take advantage of the Panoramic Survey Telescope and Rapid Response System (Pan-STARRS) Data Release 2 \citep{Flewelling2016}. We note that this survey is astrometrically tied to \textit{Gaia} \citep{Gaia_Collaboration2018}. The offset between HST and Pan-STARRS images computed using an ensemble of 375 common detections (see Fig.~\ref{offset} left panel) is comparable to the one presented in \citetalias{Franco2018}: $\Delta$RA\,=\,-96\,$\pm$\,83 mas and $\Delta$Dec\,=\,252\,$\pm$\,107 mas. 

The comparison of the positions of 69 sources in common between our 3 GHz VLA catalog (5$\sigma$ detections; Rujopakarn et al., in prep.) and Pan-STARRS within a radius of 0\farcs6 shows that there is no offset between both images (Fig.~\ref{offset} middle panel). To reduce the risk of misidentification, in all the astrometric analysis, we only retained galaxies that had been detected at least twice in the same filter during the Pan-STARRS survey. The average deviations are found to be $\Delta$RA\,=\,0\,$\pm$\,98 mas and $\Delta$Dec\,=\,12\,$\pm$\,160 mas.

Similarly, we find no offset between our ALMA sources (both the Main catalog presented in \citetalias{Franco2018} and the supplementary catalog presented in the following) and their VLA counterparts for the 27 galaxies in common between both catalogs (Fig.~\ref{offset}, right panel). The average offset is $\Delta$RA\,=\,3\,$\pm$\,113 mas and $\Delta$Dec\,=\,16\,$\pm$\,93 mas, well within the expected uncertainties for S/N $\sim$ 4 sources \citep{Ivison2007, Hatsukade2018}. 
We note here that we derive this astrometric correction using both the main catalog from \citetalias{Franco2018} and the supplementary catalog discussed in the following sections of this work. The excellent agreement in the astrometry of the VLA, ALMA, and Pan-STARRS implies that it is most likely the HST coordinate system that needs to be corrected. 

After subtracting this systematic and global offset from the HST data, the residuals offsets present spatially coherent patterns (see Fig.~\ref{Systematic_offset}). Each arrow represents the median offset between Pan-STARRS and HST positions, for a sliding median containing on average 15 sources. This local offset varies with position in the GOODS--South field, and we refer to this as a distortion offset artificially introduced during the mosaicing of the HST data. The absolute value of the distortion offset is lower than the systematic offset, but it is not negligible, and can reach values higher than 0\farcs15 at the edge of our GOODS-ALMA survey. These local distortions in the CANDELS astrometry are likely to originate from distortions in the ground-based images that were used for astrometric reference when the HST data were mosaicked.

\begin{figure}
   \centering
   \includegraphics[width=\hsize]{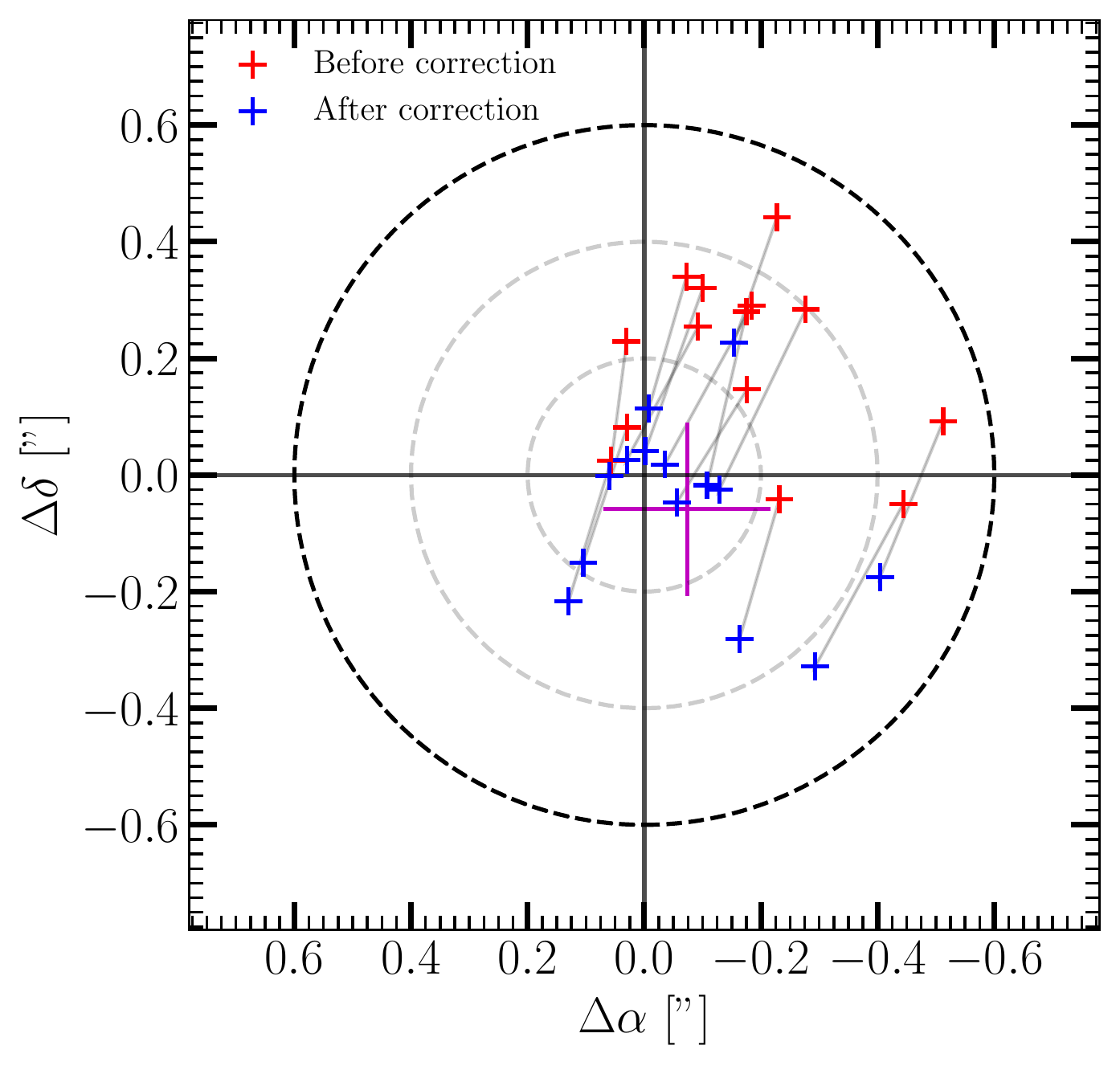}
 \caption{Positional offset (RA$_{HST}$ - RA$_\text{ALMA}$, Dec$_{HST}$ - Dec$_\text{ALMA}$) between \textit{HST} and ALMA, before (red crosses) and after (blue crosses) the correction of both a global systematic offset and a local distortion offset. The black dashed circle corresponds to the cross-matching limit radius of 0\farcs6. The gray dashed circles show positional offsets of 0\farcs2 and 0\farcs4, respectively. The mean offset and the standard deviation are shown by the magenta cross.}
         \label{positional_offset_before_after}
   \end{figure}

The combined effect of the global offset and distortion offset between the ALMA and HST positions is illustrated in Fig.~\ref{positional_offset_before_after} and listed in Table~\ref{table_offset} both before and after applying the global offset of $\Delta$RA\,=\,-96\,$\pm$\,83 mas and $\Delta$Dec\,=\,252\,$\pm$\,107 mas and the distortion offset.
With the exception of two galaxies for which the offset between the ALMA detected position and that of HST is $\sim$\,0\farcs4 (after correction of both the global offset and the distortion offsets, AGS27 and AGS34), all other galaxies have a difference in the two positions of $<$\,0\farcs33. The average deviation after correction is $-$72\,$\pm$\,143\,mas in RA and $-$58\,$\pm$\,143\,mas in Dec for the sample of galaxies selected in this study (indicated by a magenta cross in Fig.~\ref{positional_offset_before_after}).
The updated RA and Dec positions derived for all galaxies in the \cite{Guo2013} catalog after the correction of both systematic and local offset are given at \url{https://github.com/maximilienfranco/astrometry/}.

\section{ALMA Main and Supplementary Catalogs}\label{Sect::Catalog_creation}

   \begin{figure}
   \centering
   \begin{minipage}[t]{.5\textwidth}
\resizebox{\hsize}{!} {
\includegraphics[width=\hsize]{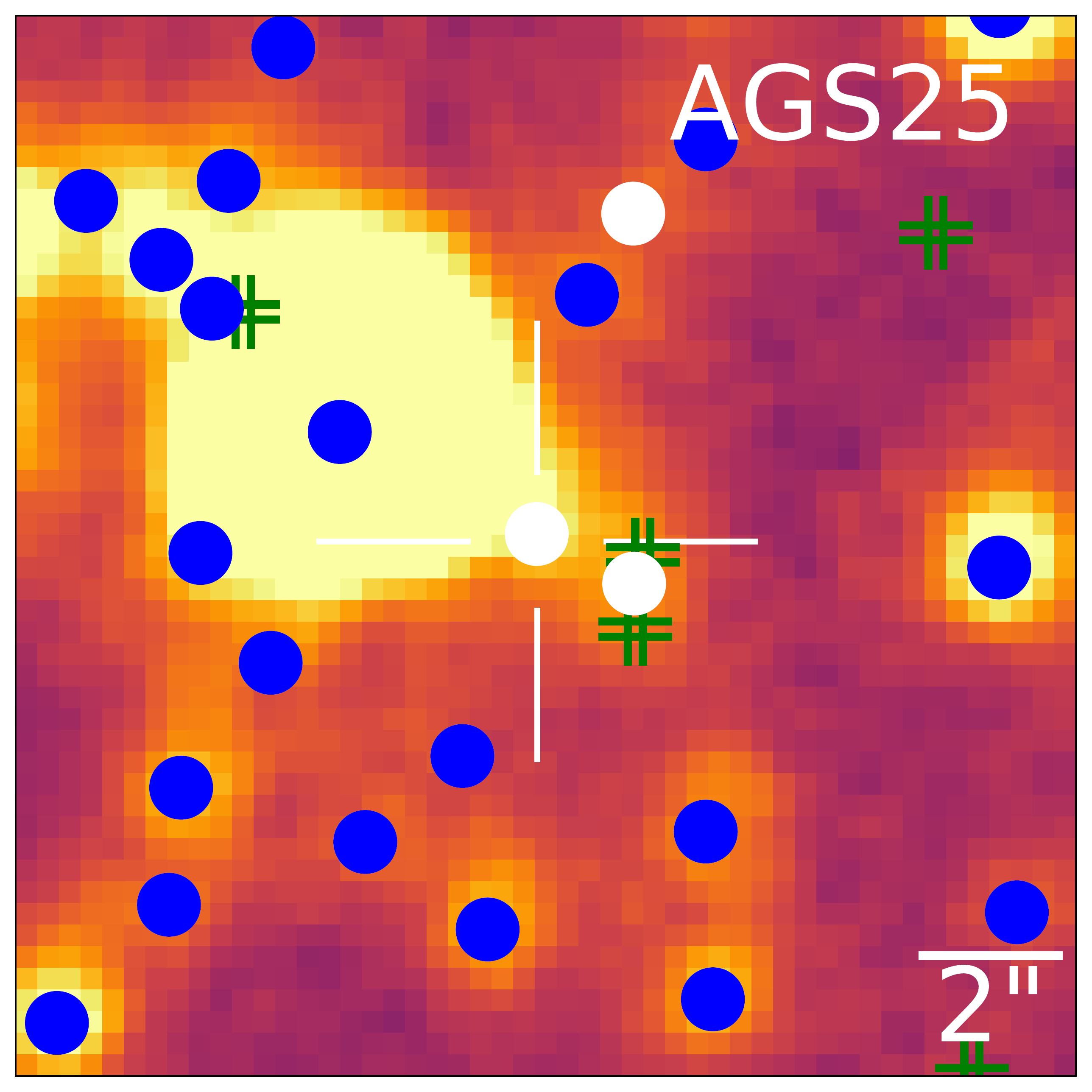}\\
\includegraphics[width=\hsize]{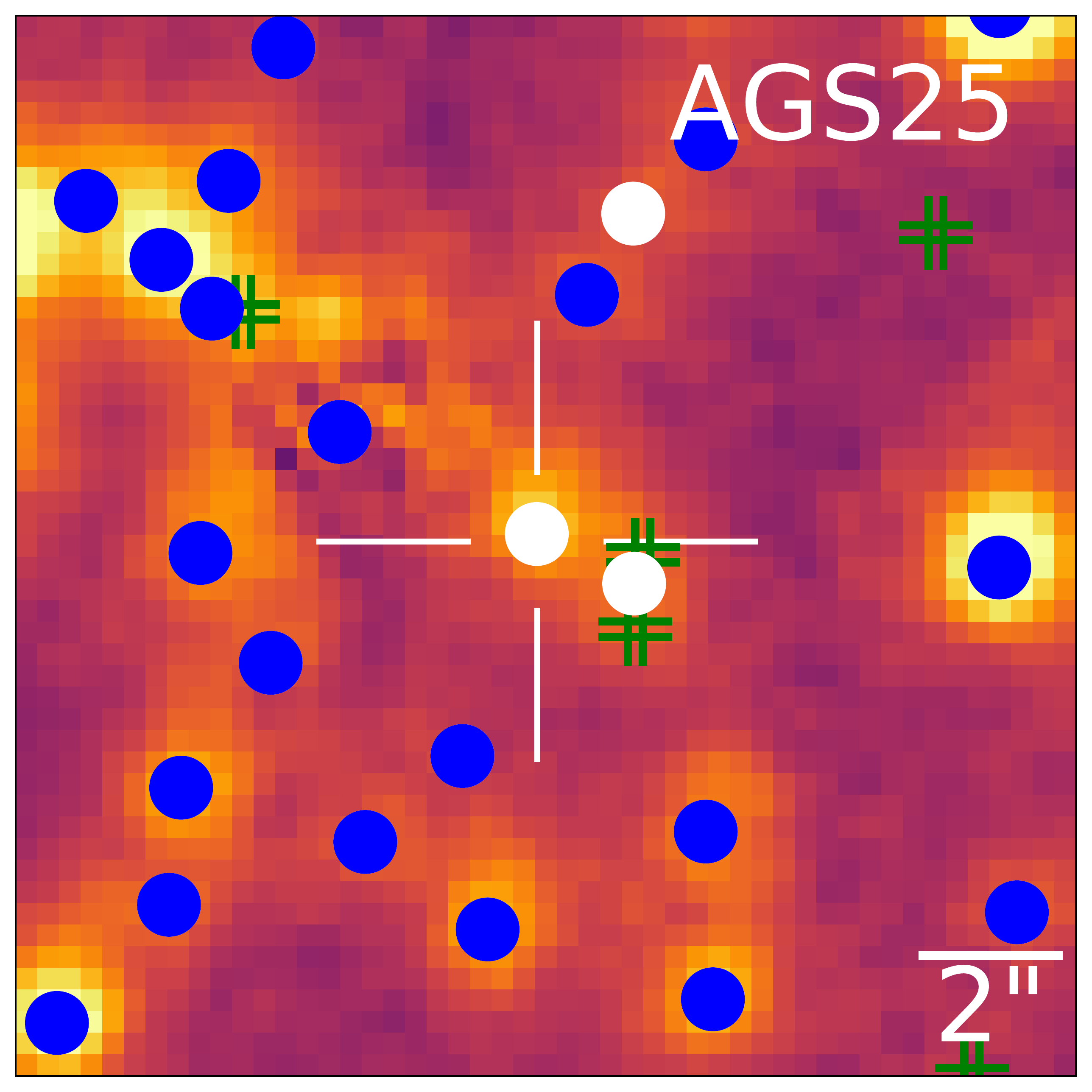}\\
}
\end{minipage}
\caption{IRAC 3.6$\mu$m image (17\arcsec\,$\times$\,17\arcsec) centered on the position of the ALMA detection. We show the image before (left panel) and after (right panel) the subtraction with \texttt{GALFIT} of the bright source ID$_{ZFOURGE}$\,=\,11024 (ID$_\text{CANDELS}$\,=\,8067) located to the northeast of the detection and which masks the emission of the source located at the ALMA position. After subtraction we can clearly see emission located in the central position which suggests that the source is not present in \cite{Ashby2015} only because of blending. Green double crosses show sources only from the GOODS--South CANDELS catalog and white circles show sources only from the ZFOURGE catalog. Blue circles show common sources to both optical catalogs (i.e., sources with an angular separation lower than 0\farcs4).} 
         \label{galfit_extraction}
   \end{figure}

\begin{figure*}
\centering
\includegraphics[width=\hsize,clip]{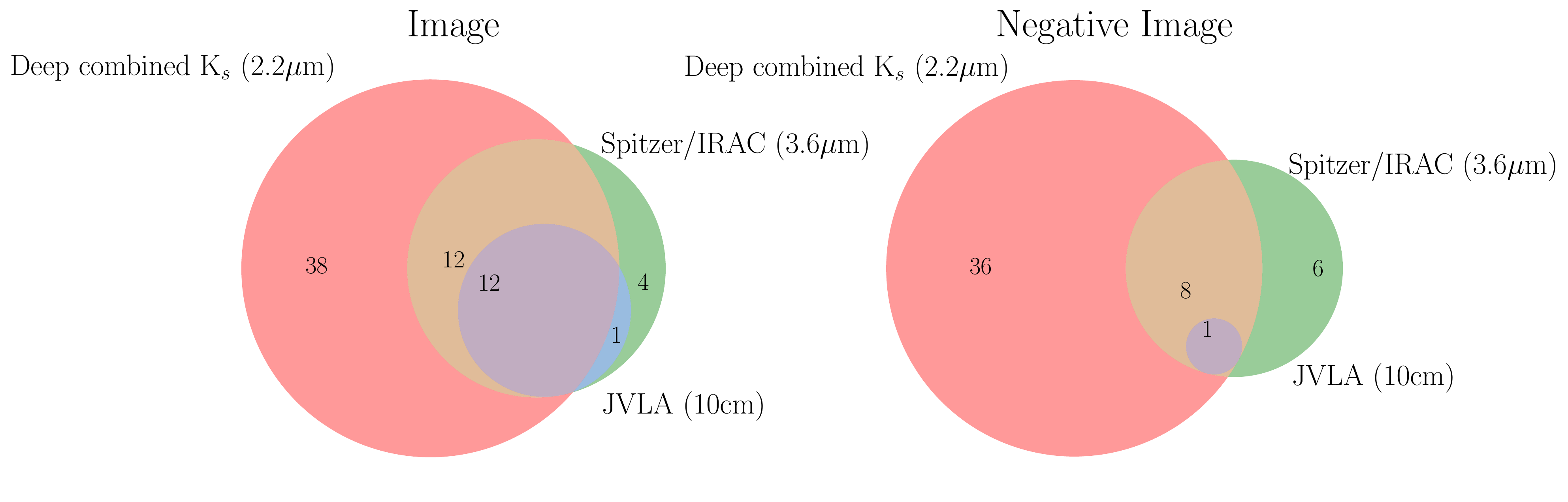}
\caption{Number of sources cross-matched between the ALMA 3.5$\sigma$ detection and the ZFOURGE \citep{Straatman2016}, S-CANDELS \citep{Ashby2015} and GOODS-VLA (PI: W.Rujopakarn, private communication) catalogs in the image (left panel) and in the negative image (right panel), within a radius of 0\farcs60 for the ZFOURGE and the VLA catalogs and 0\farcs70 for the S-CANDELS catalog. Beforehand, we previously removed from the image the sources that had been detected in \citetalias{Franco2018}. For example, in the "direct" image, among the 29 source detected both with ALMA at 3.5$\sigma$ and in the \textit{Spitzer}/IRAC channel 1 image, 24 are also detected in the ultra-deep $K_s$ image, and 13 are detected with the VLA.}
\label{cross_match}
\end{figure*}

   \begin{figure}
   \centering
\includegraphics[width=\hsize]{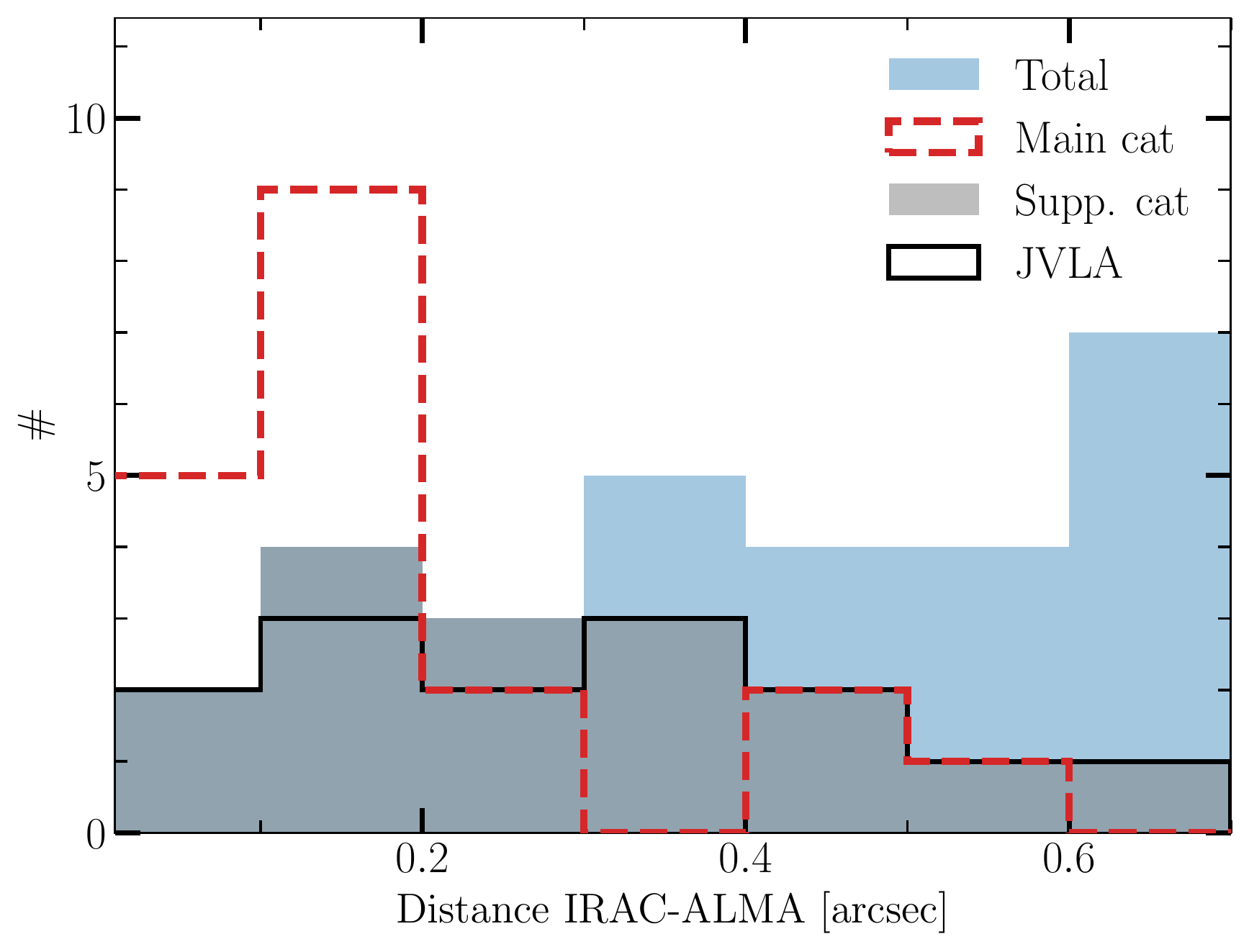}
      \caption{Distance between the ALMA positions and the closest IRAC galaxies listed in the S-CANDELS catalog for the sources presented in \citetalias{Franco2018} (red-dashed line), for the supplementary catalog (gray) and for all of the 3.5$\sigma$ detections (blue). We also represent the VLA counterparts (at 3GHz) with a solid black line. All of the ALMA 3.5$\sigma$ detections with both IRAC and VLA counterparts between 0\farcs3 and 0\farcs7 from the ALMA detections have been selected.}
         \label{Distance_IRAC_ALMA}
   \end{figure}

\subsection{ALMA Main Catalog}\label{Sect::ALMAcat}
The main ALMA catalog consists of 20 sources detected above 4.8\,$\sigma$ \citepalias{Franco2018}. This catalog was built without any prior assumptions. It is a blind source extraction down to 4.8\,$\sigma$, using \textsc{Blobcat} \citep{Hales2012}.

The detection limit was set to S/N $\geq$ 4.8 due to the large number of independent beams that leads to spurious detections that become rapidly more numerous than the number of robust detections below this threshold, despite the tapering at 0\farcs60 (see Fig.~4-left panel in \citetalias{Franco2018}). Here the 4.8\,$\sigma$-limit concerns the central pixel detection threshold ($\sigma_p$\,=\,4.8) and is associated with a constraint on the adjacent surrounding pixels that are included in the source if they pass a detection threshold of $\sigma_f$\,=\,2.7. This combination of parameters ensures an 80\,\% purity where the purity criterion $p_c$ is defined as:
\begin{equation}
p_c(>S/N)\,=\,\frac{N_p -N_n}{N_p} > 0.8 \ ,
\label{quality_criteria}
\end{equation}
\noindent where $N_p$ and $N_n$ are the number of positive and negative peaks across the whole survey at a given S/N. The negative peaks refer to the detections in the negative image (in other words detections in the continuum image multiplied by -1). The negative image, which by definition has no source signal, nevertheless preserves the correlated noise generated by the beam size, and local variations in sensitivity. This image gives a good indication of the number of spurious sources caused by an expected Gaussian noise as a function of the detection threshold. We searched for negative peaks above the same N-$\sigma$ threshold as the positive ones to determine the fraction of spurious detections.

The initial catalog coming out of the blind source extraction contains 23 detections including 3 detections that we flagged as spurious in \citetalias{Franco2018} (marked with an asterisk in the Table~2 of \citetalias{Franco2018}). Finding three spurious detections out of a total of 23 detections matches the 4\,$\pm$\,2 spurious sources expected, based on the difference between positive and negative peaks above 4.8\,$\sigma$. 
These three most-likely spurious sources are the only ALMA detections without counterparts in the ultra-deep 3.6 and 4.5\,$\mu$m IRAC images available for this field (26.5 AB mag, \citealt{Ashby2015}). The possibility that these sources are most likely spurious was later on confirmed by \cite{Cowie2018} using a 100 arcmin$^2$ survey of the field down to an RMS $\sim$0.56\,mJy at 850\,$\mu$m with SCUBA-2. None of the three detections listed as spurious were detected by SCUBA-2, while 17 out of our 20 brightest ALMA sources were detected. Moreover these three most-likely spurious sources AGS14, AGS16 and AGS19 are globally sources with a lower S/N than the others (these galaxies are classified from AGS1 to AGS20 according to their S/N).
The remaining 3 sources (AGS21, AGS22 and AGS23) were either confirmed with ALMA by \cite{Cowie2018} -- AGS21 and AGS23 -- or lie outside of the SCUBA-2 field of view -- AGS22. We note however, that AGS22 is both at the very limit of our detection threshold and not show any IRAC counterpart. In view of the very large number of expected spurious detections at the 3.5$\sigma$ limit that we plan to reach in the present paper, we will adopt the strategy to limit ourselves to the most secure ALMA detections, which exhibit a clear IRAC counterpart. As a result, we will limit the original main sample to the 19 galaxies with IRAC counterparts only, leaving aside AGS22 for consistency.

   \begin{figure*}
   \centering
   \includegraphics[width=\hsize]{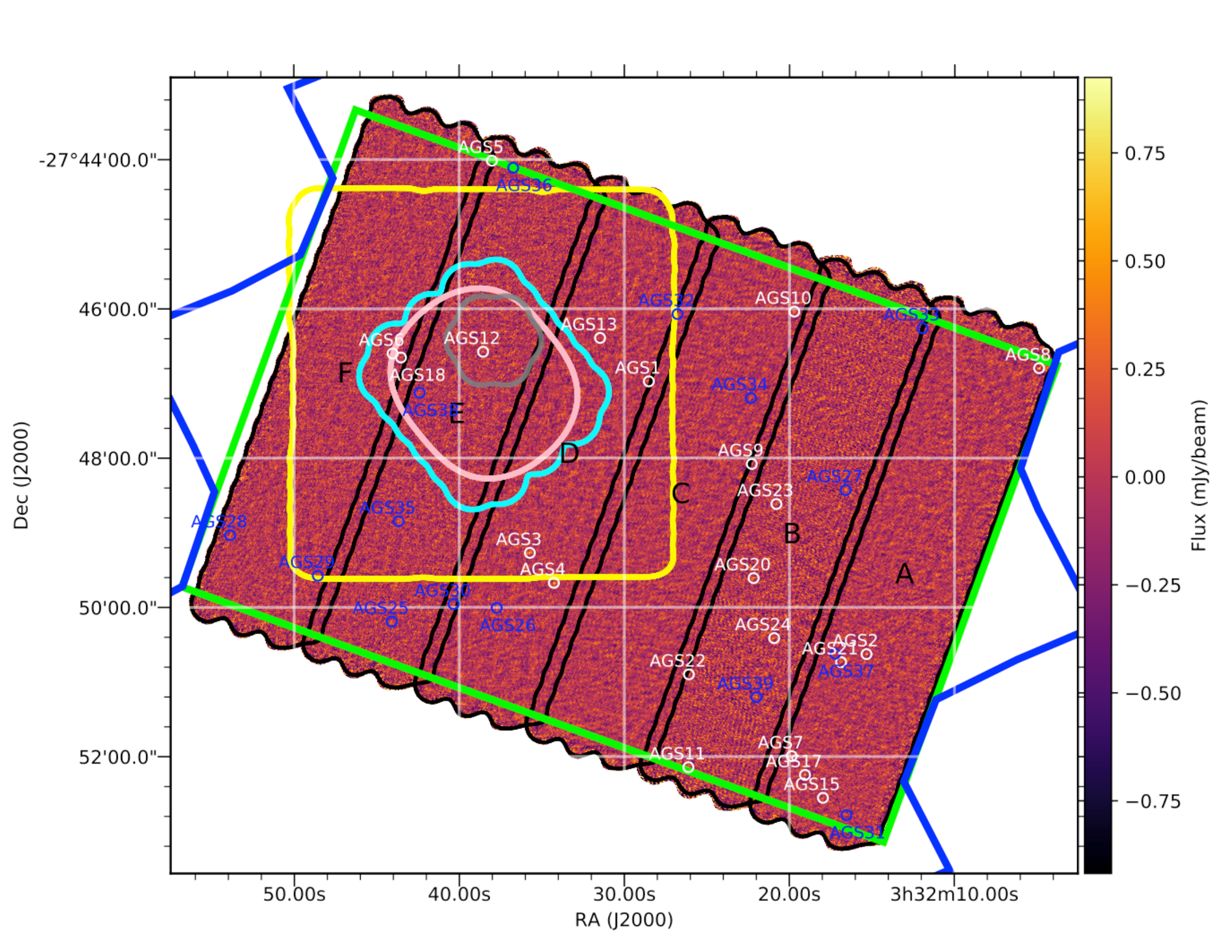}
      \caption{ALMA 1.1\,mm image tapered at 0\farcs60. The white circles have a diameter of 4 arcseconds and indicate the positions of the galaxies listed in Table~\ref{table_offset}. Black contours show the different slices (labeled A to F) used to construct the homogeneous 1.1\,mm coverage, with a median RMS\,=\,0.18\,mJy.beam$^{-1}$.  Blue lines show the limits of the HST/ACS field and green lines indicate the HST-WFC3 deep region. The cyan contours represent the limit of the \cite{Dunlop2017} survey covering all of the \textit{Hubble} Ultra Deep Field region, the yellow contours show the ASAGAO survey \citep{Hatsukade2018}, while the gray contours show the ASPECS Pilot survey \citep{Walter2016}, the pink contours show the ASPECS Large Program \citep{Decarli2019}.}
         \label{Fig::image_GOODS_South_ALMA}
   \end{figure*}

\subsection{ALMA Supplementary Catalog}\label{subSect::ALMA_Supplementary_Catalog}
\subsubsection{Using IRAC counterparts to identify robust ALMA sources down to 3.5-$\sigma$}
In the present paper, we propose to use counterparts at other wavelengths to identify robust ALMA detections below the 4.8-$\sigma$ limit of the Main Catalog described in Sect.~\ref{Sect::ALMAcat}. This approach is similar in philosophy to a prior source extraction approach, except that we start from the ALMA blind source extraction at a lower threshold and then only keep sources with counterparts already identified in the near- and mid-IR.

A total of 1077 sources are detected down to this threshold, most of which are spurious simply due to the large number of independent beams (more than one million in the 0\farcs60 tapered image). Indeed, we get an even larger number of negative peaks (N$_n$\,=\,1143) than positive ones (N$_p$\,=\,1077) which gives a negative purity.
A differential approach of Eq.~\ref{quality_criteria}, in other words, using p$_c$ (S/N -- S/N\,+\,$\Delta$ S/N) instead of p$_c$ ($>$S/N) could have allowed us to refine the prediction of the expected number of detections. However, in the S/N domain in which we work, the purity level falls (and the number of detections in the negative image increases rapidly) even using a differential approach.

We observed in \citetalias{Franco2018}  that all ALMA robust detections of the Main Catalog presented an IRAC counterpart. Hence, we impose here the requirement that all candidate detections exhibit an IRAC counterpart as well. We note that this criterion may lead us to reject real ALMA sources without any IRAC counterpart with the possible consequence of biasing our sample toward the most massive galaxies, but this is for the sake of the robustness of the sources. ALMA sources without any IRAC counterpart may well exist (see e.g., \citealt{Williams2019}) and would be particularly interesting to analyze, but this is out of the scope of the present paper. To be detected by ALMA at 1.1\,mm and missed with IRAC at the depth of this study, a star-forming galaxy should be located at high redshift ($z$\,$>$\,5) and/or extremely reddened as in the well-known case of HDF 850.1 \citep{Hughes1998,Walter2012}, even if in this case the presence of a foreground elliptical located at $\sim$1\arcsec complicates the interpretation \citep{Dunlop2004}.

We chose the cross-matching radius between the positions of the ALMA detections and the IRAC (\textit{S}-CANDELS), VLA, and ZFOURGE $K$-band catalogs to be equal to the value of the largest FWHM. It is equal to 0\farcs60, the FWHM of the tapered 1.1\,mm ALMA image, for the cross-match with the VLA and ZFOURGE $K$-band catalogs. For IRAC, although this value would be 1\farcs95 for the cross-match with the IRAC channel 1 (3.6\,$\mu$m) catalog (FWHM$_{IRAC}$\,=\,1\farcs95 at 3.6\,$\mu m $ and FWHM$_{IRAC}$\,=\,2\farcs05 at 4.5 $\mu m $), we have only considered sources closer to 0.7" (see explanation below). Before performing this cross-matching, we apply the astrometric correction to the CANDELS and ZFOURGE catalogs which use the HST reference frame, as described in Sect.~\ref{Sect::Astrometric_correction}. In order to ensure reliable positioning of the ALMA sources, we injected a realistic source distribution in the image and using the same detection techniques as described in the previous section, we quantified the differences between the injected sources and the recovered sources. After injecting $\sim$ 250\,000 sources, the spatial difference between injected and detected sources is less than 0\farcs2 for 85\% of the detected sources with a S/N between 3.5 and 4.8.

When we restrict the search area to the close neighborhood of IRAC sources (within this search radius of 0\farcs7), we get a much improved purity of 48\% with 15 negative peaks (N$_n$, right side of Fig.~\ref{cross_match}: 8+1+6) and 29 positive peaks (N$_p$, left side of Fig.~\ref{cross_match}: 12+12+4+1). This gives an estimated 14 expected real detections. Although the purity does improve a lot that way, we still need to perform a second selection step to identify the real candidates as discussed below.

In the process of cross-matching the ALMA and IRAC images, we identified, thanks to VLA images, two IRAC sources (AGS24 and AGS25) not listed in the S-CANDELS catalog (see Sect.~\ref{Sect::HST-dark}) due to blending with a bright source.
 We therefore implemented a new source extraction for those sources taking care to model the neighboring sources to obtain a clean de-blending of the IRAC sources (see Fig.~\ref{galfit_extraction}). The use of the HST catalog in band $H$ did not provide any reliable additional sources (see below).

A total of 67 sources detected with ALMA between 3.5 and 4.8\,$\sigma$ at 1.1\,mm have a counterpart in at least one of the three catalogs (see Fig.~\ref{cross_match}). We have included in this figure the two sources (AGS24 and AGS25) missed in the S-CANDELS catalog due to the presence of a bright neighbor. In comparison, in the negative image, there are 51 detections above 3.5-$\sigma$ that also fulfill these criteria. 

We note that 84\,\% (16 out of 19 sources) of the ALMA sources in the Main Catalog described in \citetalias{Franco2018}  have an IRAC counterpart closer than 0\farcs3 (red dashed line in Fig.~\ref{Distance_IRAC_ALMA}). Since our goal here is not to include all possible ALMA sources but to limit the Supplementary Catalog to the most robust candidates, we decided to impose a more stringent constraint on the association with IRAC counterparts by keeping as robust candidates those within a distance of 0\farcs3. In total, 9 ALMA sources detected between 3.5$\sigma$ and 4.8$\sigma$ fulfill this criterion. 

Looking at the remaining three sources in the Main Catalog (sources with a distance greater than 0\farcs3 from their IRAC counterpart), we see another interesting characteristic. They are all closer than 0\farcs7 from their IRAC counterpart and nearly all exhibit a radio counterpart as well (2 out of 3). In fact, out of the 19 sources in the Main Catalog, 16 (84\,\%) exhibit a radio counterpart. We therefore list in the Supplementary Catalog the sources that have both an IRAC and a radio counterpart closer than 0\farcs7. This extra condition adds an extra 6 ALMA sources detected between 3.5$\sigma$ and 4.8$\sigma$. In total, we end up with a list of 16 sources that fulfill the criteria of having an IRAC counterpart either \textit{(i)} closer than 0\farcs3 or \textit{(ii)} closer than 0\farcs7 but associated with a radio counterpart in the 3 GHz catalog.

It is possible that using these criteria does not allow us to detect all "real" ALMA detections with a S/N $>$ 3.5 but these conservative criteria ensure a high purity rate. Applying these criteria to the negative image, we find only four resulting sources. We performed Monte Carlo simulations to estimate the probability that an ALMA detection lies randomly close to a galaxy listed in the S-CANDELS catalog. We randomly define a position within GOODS--South and then measure the distance to its closest IRAC neighbor using the source positions listed in \cite{Ashby2015} for IRAC sources with S/N\,$>$\,5. We repeat this procedure 100\,000 times. This method gives results comparable to those presented in \cite{Lilly1999}. The distance from the nearest IRAC galaxies is given Table~\ref{table_offset}. In our supplementary catalog, for the farthest source to an IRAC counterpart (AGS27; 0\farcs64), the percentage of a random IRAC association is 1.38\,\%. With the exception of one other source (AGS29), the other detections have a probability of random IRAC association $\le$ 1\%.
The surface density of radio sources is significantly lower than the surface density of IRAC sources (about 8 times lower), so the probability of a false association is also lower. A radio source at a distance of 0\farcs6 from the ALMA counterpart has a probability of being a random association of $\sim$\,0.1\%. This percentage naturally depends on the depth of the catalog chosen. Using the \cite{Guo2013} catalog, we find a probability of a random association of $\sim$8\% in $H$ band at a distance of 0\farcs6 from the ALMA source, and $\sim$5\% in $K$-band using the \cite{Straatman2016} catalog.

We checked whether deeper surveys covering parts of the GOODS--South field could be used to validate or invalidate those ALMA Supplementary Catalog sources. The HUDF \citep{Dunlop2017}, the ASAGAO \citep{Hatsukade2018} and the ASPECS Large Program \citep{Decarli2019, Gonzalez_Lopez2020} surveys reach a depth of RMS $\simeq$ 35\,$\mu$\,Jy at 1.3\,mm, RMS $\simeq$ 61\,$\mu$\,Jy at 1.2\,mm, and RMS $\simeq$ 9.6\,$\mu$\,Jy at 1.2\,mm, respectively. Using the same scaling factors as those presented in \citetalias{Franco2018}, these depths correspond to RMS $\simeq$ 52\,$\mu$Jy and RMS $\simeq$\,79\,$\mu$Jy, respectively, at the wavelength of 1.1\,mm relevant to this GOODS-ALMA survey. Only three ALMA 1.1\,mm sources from the Supplementary Catalog fall in the area covered by these deeper surveys and all of them were detected and listed in the associated catalogs (see Fig.~\ref{Fig::image_GOODS_South_ALMA}), demonstrating the robustness of our approach. 
AGS29 and AGS35 are listed as sources 18 and 26, respectively, in the ASAGAO survey \citep{Hatsukade2018} while AGS38 is known as UDF16 in the HUDF survey \citep{Dunlop2017} and C15 in \cite{Gonzalez_Lopez2020}. Our independent identification of sources down to the 3.5-$\sigma$ level did not therefore introduce any spurious sources without counterparts in deeper ALMA surveys. This ensures a high degree of purity in our sample. Moreover, in addition to the completeness analysis presented in \citetalias{Franco2018}, we investigate the relative completeness of our sample compared to the ASAGAO sample. We used the sample of 45 sources presented in \cite{Hatsukade2018} in which we removed 3 sources (6, 20 and 21) located outside or on the edge of GOODS-ALMA. Considering simultaneously the sources presented in \citetalias{Franco2018} and those of this analysis, we find 7 common sources. By taking the 1\,mm peak flux measurements given in Table~3 of \cite{Hatsukade2018}, we have computed the relative completeness of GOODS-ALMA compared to ASAGAO. This is 100\% for sources with a peak flux greater than 0.6\,mJy beam$^{-1}$ (5/5) and drops sharply for lower fluxes for example 65\% (7/12) at 0.5\,mJy beam$^{-1}$.

 \subsubsection{Supplementary Catalog: Optically dark galaxies}\label{Sect::HST-dark}

As discussed above, thanks to the known position of the VLA detections, we have de-blended the IRAC sources to find two additional sources that satisfy our selection criteria (AGS24 and AGS25). Although AGS25 is listed in the ZFOURGE catalog, AGS24 is not. More interestingly, neither of these two sources have been detected by the HST even in the 1.6\,$\mu$m $H$-band (down to the 5$\sigma$ limiting depth of $H$\,=\,28 AB for a point-like source), hence they are optically dark like four sources listed in \citetalias{Franco2018}  and as also discussed in \cite{Wang2016, Elbaz2017, Schreiber2018, Yamaguchi2019,Wang2019}. In this section, we will describe these two detections.

The source AGS24 exhibits extended IRAC emission (after de-blending) that is 0$\farcs$28 away from the ALMA position. This source is also detected in the radio at 6 GHz (3.7$\sigma$) and 3 GHz (5.7$\sigma$). The S/N of this source is higher in the 0\farcs29 mosaic than in the 0\farcs60 tapered image, which suggests that it is particularly compact at 1.1\,mm. This galaxy will be presented in detail in \cite{Zhou2020}, where a stellar mass and redshift are estimated to be $z$\,$\sim$\,3.5 and M$_{\star}$\,=\,2.09$_{-0.74}^{+0.10}$ $\times$10$^{11}$M$_\odot$.

The source AGS25 is 0$\farcs$1 away from its $K$-band counterpart in the ZFOURGE catalog (after applying the astrometric correction to the position of the ZFOURGE source), the source ID$_{ZFOURGE}$\,=\,11353 with a magnitude $K$\,=\,25.9 AB shown by a circle in Fig.~\ref{galfit_extraction}. This source is not listed in the CANDELS catalog \citep{Guo2013}, nor in the S-CANDELS catalog \citep{Ashby2015}. It is marginally detected in radio at 5 and 10\,cm with a S/N ratio close to 3.7. AGS25 is close (3$\arcsec$) to a massive galaxy listed in CANDELS, ID$_\text{CANDELS}$\,=\,8067 with a stellar mass of M$_{\star}$\,=\,5.6$\times$10$^{10}$ M$_\odot$ at a redshift of $z_{sp}$\,=\,1.038). ID$_\text{CANDELS}$\,=\,8067 is the bright neighbor that explains the absence of AGS25 in S-CANDELS. We subtracted it from the IRAC image by modeling a S\'{e}rsic profile with \texttt{GALFIT} \citep{Peng2010} and measured an IRAC flux density of 0.81$\pm$0.19\,mJy (see Fig.~\ref{galfit_extraction}). The IRAC source is 0$\farcs$43 away from the ALMA position of AGS25. 

Since the ALMA source is only 0$\farcs$1 away from the ZFOURGE source ID$_{ZFOURGE}$\,=\,11353, we use the stellar mass and redshift from the ZFOURGE catalog for this source. The characteristics of this galaxy make it particularly interesting, with z$_{AGS25}$\,=\,4.64$_{-0.26}^{+0.25}$ and M$_{\star,AGS25}$\,=\,2.5\,$\times$\,10$^{10}$M$_\odot$. 

We note that the six optically dark galaxies discovered with this survey (four described in \citetalias{Franco2018} and two described in this paper) appear to be particularly distant ($z$\,$\ge$\,3.5). Among these galaxies we also see that two are particularly massive (AGS4 and AGS24) with M$_\star$\,$>$\,10$^{11}$M$_\odot$.

Ultimately, we end up with a list of 16 sources in the Supplementary Catalog including the two optically dark sources AGS24 and AGS25. Their properties are listed in Table~\ref{table_offset} and Table~\ref{general_properties}.

\subsection{Consistency test of the Supplementary Catalog: Stellar mass distribution}
   \begin{figure}
   \centering
   \includegraphics[width=\hsize]{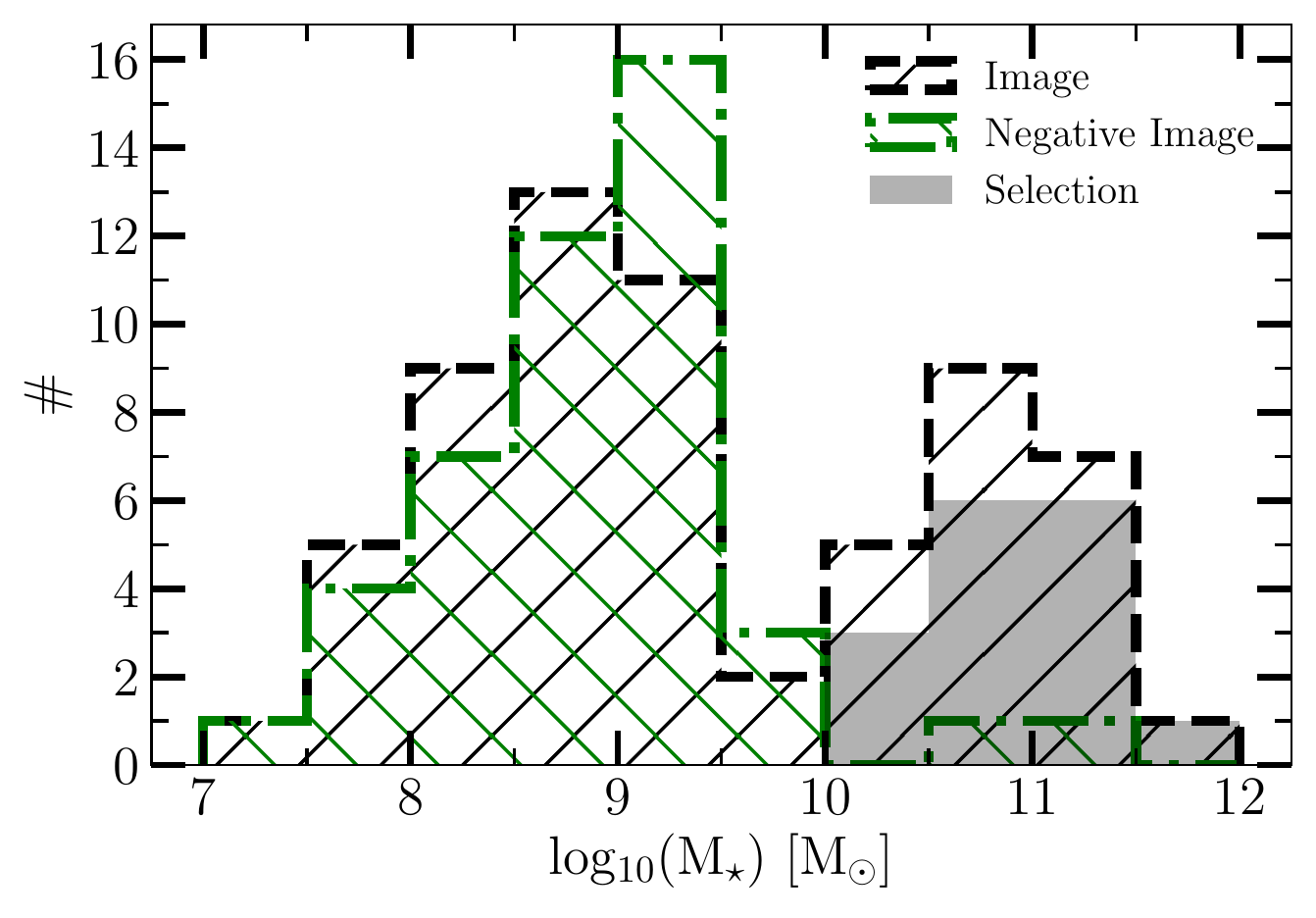}
      \caption{Stellar mass distribution resulting from the cross-matched between the ZFOURGE catalog and the detection at 3.5$\sigma$ in the image (dashed black line) and in the negative image (dashed-dotted green lines). The selection work presented in this study allowed us to select 16 galaxies (in gray) all located in the second peak of the stellar mass distribution (absent in the negative image).}
         \label{Mass_histogramm}
   \end{figure}

If we compare the nature of the counterparts of ALMA detections above 3.5-$\sigma$ in the image and the negative image, we can see a difference that strengthens the robustness of the 16 sources of the Supplementary Catalog. There are 62 and 45 detections above 3.5-$\sigma$ in the ALMA image that have a counterpart in the $K$-band of the ZFOURGE catalog in the image and negative image, respectively. The stellar mass distributions of both samples are represented by dashed black and green lines for the image and negative image, respectively, in Fig.~\ref{Mass_histogramm}. Both histograms show the same behavior at stellar masses below 10$^{10}$\,M$_\odot$ but there are nearly no galaxies (2) above this mass threshold in the negative image while there is a second bump in the histogram of the sources in the real image. Massive galaxies being rarer than low mass ones, the probability to get an association with such galaxy is lower and the fact that there is a second bump at high stellar masses in the real image supports the conclusion that these may be real sources. The Supplementary Catalog histogram shown in filled gray matches very nicely this second bump of massive galaxies. We recall that we did not impose any criterion of brightness or stellar mass in the selection of the Supplementary Catalog but only distances to IRAC, $K$-band and radio sources. If we limit ourselves to the galaxies above a stellar mass of 10$^{10}$M$_\odot$, we can see that the number of sources in the Supplementary Catalog is close to the difference in the number of detections between the image and negative images. 

In the sample of cross-matched galaxies from the positive image, 22/63 galaxies ($\sim$ 35\,\%) have a stellar mass greater than 10$^{10}$M$_\odot $, compared with only 2/45 galaxies ($\sim$4\,\%) in the negative image (see Fig.~\ref{Mass_histogramm}). A Kolmogorov-Smirnov (KS) test on these data gives a p-value of 1.2$\times$10$^{-2}$ between these two samples, meaning that the likelihood that the two samples were drawn from the same distribution is unlikely (can be rejected at 3-$\sigma$ confidence).

When we remove the sample of 16 galaxies listed in Table~\ref{table_offset}, the two samples become more similar. The p-value from a KS test then reaches 0.88. This means that once the galaxies in our study have been removed, the detections that remain have as high a probability of originating from the same parent sample as the negative image detections, so that they are no longer statistically different from noise.

This suggests that not only the Supplementary Catalog is robust but also that there is little margin for an extra population of real sources that we would have missed.

\begin{table*}\footnotesize     
\centering          
\begin{tabular}{r c c c c c c c c c c}   
\hline    
ID&RA$_{\rm ALMA}$ &Dec$_{\rm ALMA}$ &RA$_{HST}$ &Dec$_{HST}$ & $\Delta _{HST_1} $ &$\Delta _{HST_2} $  &$(\Delta\alpha)_{HST}$ &$(\Delta\delta)_{HST}$ &$\Delta_{IRAC}$ & \%RaA \\
 &$\deg$  &$\deg$&$\deg$ &$\deg$ &arcsec&arcsec&arcsec&arcsec&arcsec&\%\\
(1)&(2)&(3)&(4)&(5)&(6)&(7)&(8)&(9)&(10)&(11)\\
\hline    
\hline       
AGS24 &  53.087178 & -27.840217 &        ... &        ... &     ... &     ... &      ... &     ...  &    0.28 & 0.28 \\
AGS25 &  53.183710 & -27.836515 &        ... &        ... &     ... &     ... &      ... &     ...  &    0.43 & 0.63 \\
AGS26 &  53.157229 & -27.833468 &  53.157238 & -27.833446 &    0.09 &    0.18 &    0.075 &   -0.232 &    0.31 & 0.33 \\
AGS27 &  53.069132 & -27.807155 &  53.068992 & -27.807169 &    0.45 &    0.44 &    0.151 &   -0.279 &    0.64 & 1.38 \\
AGS28 &  53.224467 & -27.817214 &  53.224476 & -27.817151 &    0.23 &    0.06 &    0.029 &   -0.231 &    0.09 & 0.03 \\
AGS29 &  53.202362 & -27.826284 &  53.202340 & -27.826190 &    0.35 &    0.11 &    0.065 &   -0.226 &    0.59 & 1.14 \\
AGS30 &  53.168097 & -27.832632 &  53.168025 & -27.832509 &    0.50 &    0.27 &    0.074 &   -0.215 &    0.26 & 0.24 \\
AGS31 &  53.068906 & -27.879739 &  53.068851 & -27.879698 &    0.23 &    0.07 &    0.120 &   -0.194 &    0.13 & 0.07 \\
AGS32 &  53.111595 & -27.767860 &  53.111564 & -27.767771 &    0.34 &    0.04 &    0.099 &   -0.280 &    0.14 & 0.07 \\
AGS33 &  53.049749 & -27.771007 &  53.049662 & -27.770929 &    0.40 &    0.13 &    0.148 &   -0.310 &    0.21 & 0.14 \\
AGS34 &  53.093099 & -27.786607 &  53.092938 & -27.786582 &    0.52 &    0.44 &    0.108 &   -0.267 &    0.36 & 0.44 \\
AGS35 &  53.181971 & -27.814127 &  53.181989 & -27.814120 &    0.06 &    0.25 &    0.073 &   -0.241 &    0.12 & 0.05 \\
AGS36 &  53.153025 & -27.735192 &  53.152971 & -27.735114 &    0.33 &    0.11 &    0.068 &   -0.298 &    0.37 & 0.47 \\
AGS37 &  53.071752 & -27.843712 &  53.071694 & -27.843631 &    0.34 &    0.04 &    0.149 &   -0.273 &    0.01 & 0.00 \\
AGS38 &  53.176650 & -27.785435 &  53.176577 & -27.785446 &    0.24 &    0.33 &    0.068 &   -0.240 &    0.40 & 0.55 \\
AGS39 &  53.091634 & -27.853413 &  53.091606 & -27.853342 &    0.27 &    0.04 &    0.122 &   -0.228 &    0.11 & 0.05 \\
\hline
\end{tabular}
\caption{Details of the positional differences between ALMA and \textit{HST}-WFC3 for our catalog of galaxies identified in the 1.1\,mm-continuum image. Columns: (1) Source ID; (2), (3) Coordinates of the detections in the ALMA image (J2000);  (4), (5) Positions of \textit{HST}-WFC3 $H$-band counterparts when applicable from \cite{Guo2013}, (6), (7) Distances between the ALMA and \textit{HST} source positions \textit{before} ($\Delta_{HST_1}$) and \textit{after} ($\Delta_{HST_2}$) applying both the systematic and local offset correction presented in Sect.~\ref{Sect::Astrometric_correction};  (8), (9) Offset to be applied to the \textit{HST} source positions, which includes both the global systematic offset and the local offset; (10) Distance from the closest IRAC galaxy; (11) IRAC random association (RaA) between the ALMA detection and the closest IRAC galaxy.}
\label{table_offset}  
\end{table*}

\begin{table*}\footnotesize     
\centering          
\begin{tabular}{c c c c c c c c c c c c}      
\hline 
ID ALMA  & ID$_{CLS}$ & ID$_{ZF}$ & ID$_{S-CLS}$ & z$_{CLS}$ & $z_{ZF}$& $z_{sp}$ & S/N & Flux &log$_{10}(\text{M}_*)$  & S$_{3GHz}$\\
&&&&&&&& mJy &M$_\odot$ &$\mu$Jy\\
(1)&(2)&(3)&(4)&(5)&(6)&(7)&(8)&(9)&(10)&(11)\\
\hline
\hline
    AGS24  &  ...  &  ...  &       ...                  &   ... & ...   & 3.472$^{sp}$ & 3.93  &  0.88\,$\pm$\,0.22$^{pf}$ & 11.32\dag &  12.43\,$\pm$\,2.19 \\ 
    AGS25  &  ...  & 11353 &       ...                  &   ... & 4.644 &...           & 4.36  &  0.81\,$\pm$\,0.19$^{pf}$ & 10.39     &   6.69\,$\pm$\,1.78 \\
    AGS26  &  8409 & 11442 &       J033237.75-275000.8  & 1.711 & 1.592 & 1.619$^{sp}$ & 4.31  &  0.97\,$\pm$\,0.15\ \ \ \ & 10.91     &  85.09\,$\pm$\,2.55 \\
    AGS27  & 11287 & 14926 &       J033216.54-274825.7  & 4.931 & 4.729 &...           & 3.76  &  1.43\,$\pm$\,0.28\ \ \ \ & 10.93     &   5.95\,$\pm$\,1.86 \\
    AGS28  & 10286 & 13388 &       J033253.87-274901.9  & 2.021 & 2.149 &...           & 4.10  &  1.56\,$\pm$\,0.21\ \ \ \ & 11.17     &  17.19\,$\pm$\,1.85 \\
    AGS29  &  9242 & 12438 &       J033248.53-274934.8  & 1.346 & 1.071 & 1.117$^{sp}$ & 3.56  &  0.61\,$\pm$\,0.18$^{pf}$ & 10.77     &  65.01\,$\pm$\,2.38 \\
    AGS30  &  8557 & 11581 &       J033240.33-274957.3  & 0.646 & 0.672 & 0.65$^{sp}$  & 4.00  &  0.67\,$\pm$\,0.17$^{pf}$ & 10.40     &        ...          \\
    AGS31  &  3584 &  6153 &       J033216.53-275247.0  & 2.686 & 2.445 &...           & 3.93  &  0.72\,$\pm$\,0.19$^{pf}$ & 11.38     &        ...          \\
    AGS32  & 16822 & 19964 &       J033226.78-274604.2  & 4.526 & 4.729 &...           & 3.92  &  1.23\,$\pm$\,0.16\ \ \ \ & 11.00     &   4.47\,$\pm$\,1.38 \\
    AGS33  & 16558 & 19463 &       J033211.93-274615.5  & 2.571 & 2.676 &...           & 3.85  &  1.77\,$\pm$\,0.27\ \ \ \ & 10.71     &  21.20\,$\pm$\,2.84 \\
    AGS34  & 14035 & 17374 &       J033222.32-274711.9  & 2.866 & 2.750 &...           & 3.72  &  0.55\,$\pm$\,0.15$^{pf}$ & 10.82     &  15.55\,$\pm$\,1.98 \\
    AGS35  & 10497 & 14146 &       J033243.67-274851.0  & 2.986 & 9.476 &...           & 3.71  &  1.16\,$\pm$\,0.21\ \ \ \ & 10.85     &  31.49\,$\pm$\,1.42 \\
    AGS36  & 20859 & 23463 &       J033236.70-274406.6  & 0.646 & 0.663 & 0.665$^{sp}$ & 3.66  &  0.74\,$\pm$\,0.21$^{pf}$ & 10.46     &  11.71\,$\pm$\,1.60 \\
    AGS37  &  7184 & 10241 &       J033217.22-275037.3  & 1.971 & 1.864 & 1.956$^{sp}$ & 3.64  &  1.10\,$\pm$\,0.16\ \ \ \ & 11.19     &  22.61\,$\pm$\,4.39 \\
    AGS38  & 14638 & 17465 &       J033242.37-274707.8  & 1.346 & 1.323 & 1.314$^{sp}$ & 3.62  &  1.00\,$\pm$\,0.16\ \ \ \ & 11.08     &   9.92\,$\pm$\,2.28 \\
    AGS39  &  6131 &  9248 &       J033222.00-275112.3  & 2.906 & 2.360 &...           & 3.62  &  0.80\,$\pm$\,0.23$^{pf}$ & 10.57     &  17.24\,$\pm$\,2.29 \\
\hline                  
\end{tabular}
\caption{Characteristics of the sources forming the Supplementary Catalog. Columns: (1) Source ID; (2), (3), (4) IDs of the HST-WFC3 (from the CANDELS catalog), ZFOURGE and IRAC (SEDS catalog) counterparts of these detections; (5), (6) Photometric redshifts from the CANDELS catalog \citep{Guo2013}, z$_{CLS}$, and ZFOURGE catalog \citep{Straatman2016}, z$_{ZF}$ (we note that AGS24 has a redshift of $z$$\simeq$3.472 determined by \cite{Zhou2020}, see Sect.~\ref{Sect::HST-dark}) ; (7) Spectroscopic redshift when available (see Sect.~\ref{Sect::Redshifts}), flagged with an "sp" exponent to avoid confusion; (8) S/N of the detections in the 0\farcs60 mosaic. This S/N is given for the peak flux; (9) Flux and error on the flux as explained in Sect.~\ref{Flux_and_size_measurement}. $^{pf}$ indicates that the flux used is the peak flux, as the size measured by \texttt{uvmodelfit} is below the size limit given by Eq.~\ref{Marti}. If there is no indication, the flux used is the flux given by \texttt{uvmodelfit}; (10) Stellar mass as described in Sect.~\ref{Sect::Redshifts}; (11) 3GHz flux density from VLA (PI W.Rujopakarn, private communication).} 
\label{general_properties}  
\end{table*}

\section{Catalog}\label{Sect::Catalog}
The positions of the ALMA sources listed in the Main and Supplementary catalogs are shown in Fig.~\ref{Fig::image_GOODS_South_ALMA} where they can be compared to the locations of other ALMA surveys. The postage-stamp images of the sources are shown in Appendix~\ref{multiwav_main_cat}.

\subsection{Redshifts and stellar masses}\label{Sect::Redshifts}

\begin{figure*}
\centering
\begin{minipage}[t]{1.\textwidth}
\resizebox{\hsize}{!} {
   \includegraphics[width=0.46\hsize]{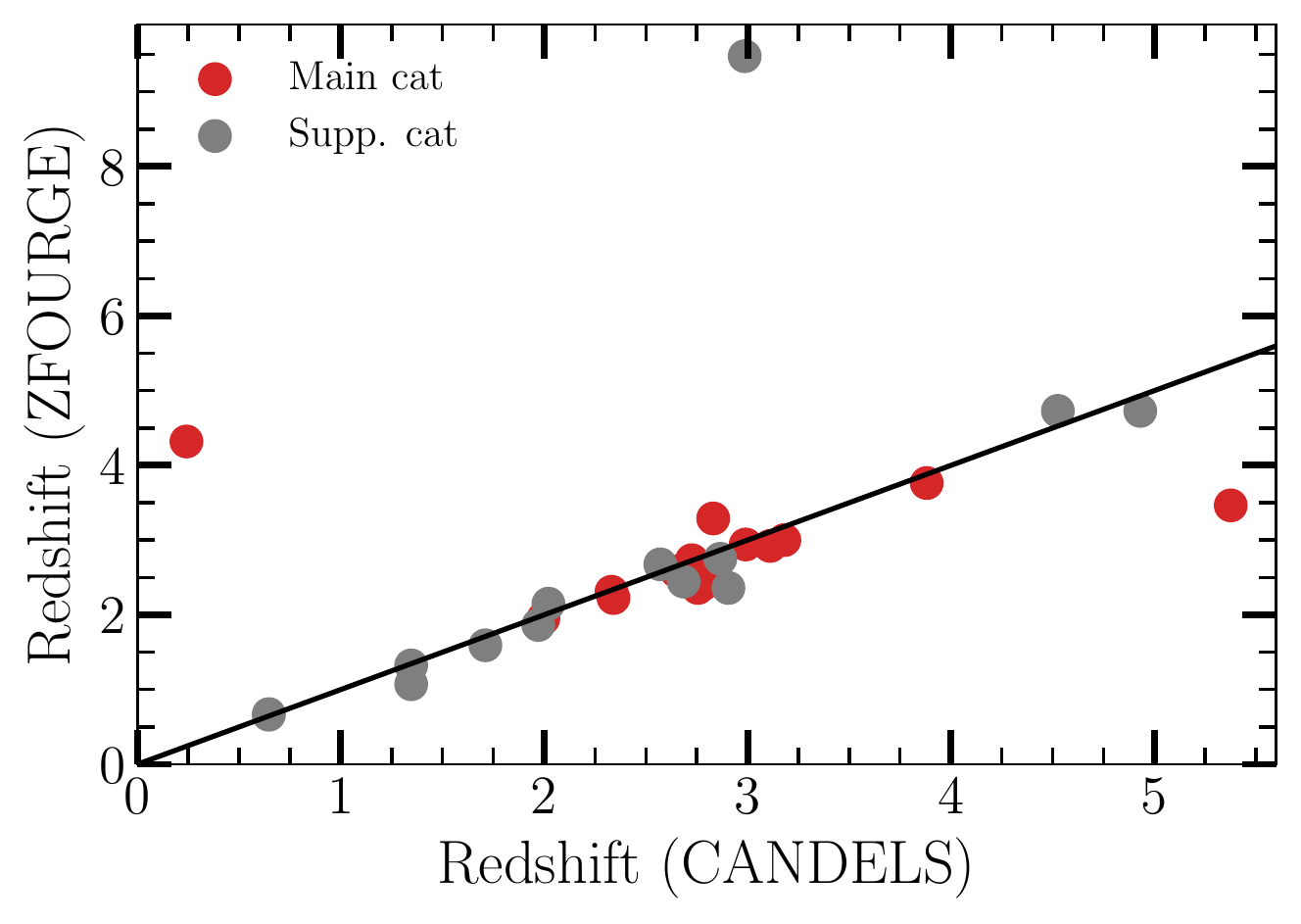}
      \includegraphics[width=0.5\hsize]{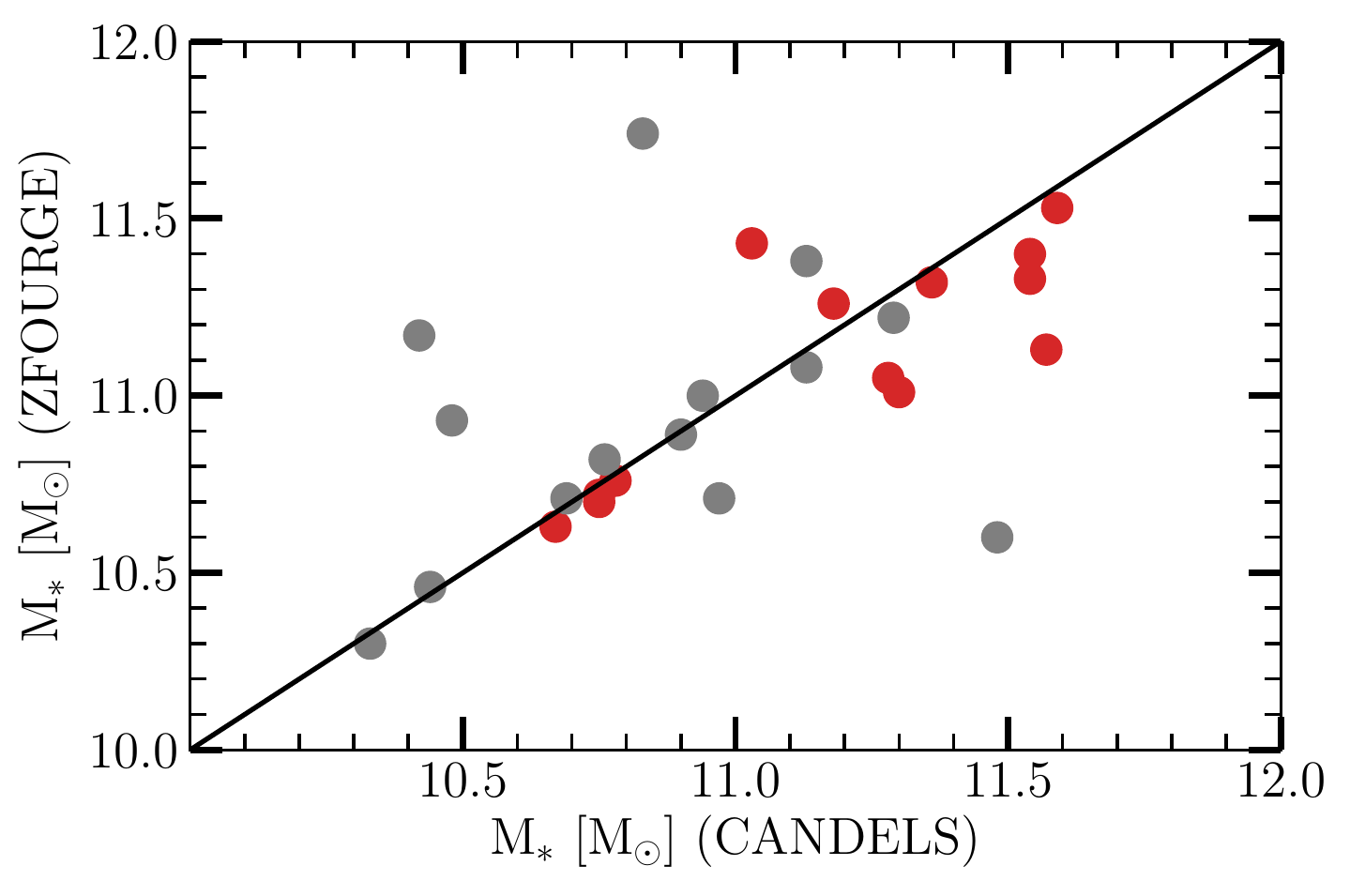}
      }
\end{minipage}
\caption{Comparison of redshift (left panel) and stellar mass (right panel) from the CANDELS and the ZFOURGE catalogs. Solid black lines indicate $z_\text{ZFOURGE}$\,=\,$z_\text{CANDELS}$. The galaxies presented in \citetalias{Franco2018} (Main Catalog) are shown in red, while the galaxies presented in this paper are shown in gray. The stellar mass has been scaled from a Chabrier IMF to a Salpeter IMF by applying a factor of 1.7 in the ZFOURGE catalog.  In this paper, we will take, with the exception of AGS35 which has an inconsistent redshift ($z_{AGS35,ZFOURGE}$\,=\,9.47 and for which we will take the data from the CANDELS catalog), the redshifts and stellar masses from the ZFOURGE catalog. The CANDELS stellar masses come from \cite{Santini2015}.}
\label{compare_z}
\end{figure*}

Except for the two optically dark galaxies, AGS24 and AGS25 (discussed in Sect.~\ref{Sect::HST-dark}), all sources listed in the ALMA Supplementary Catalog have a photometric redshift reported in the CANDELS \citep{Guo2013} and ZFOURGE \citep{Straatman2016} catalogs. Photometric redshifts inferred by these different teams are listed in Col.(5) and Col.(6) of Table~\ref{general_properties}. They are in excellent agreement (see Fig.~\ref{compare_z}, left) with $<$\,$|z_\text{CANDELS}$ - $z_\text{ZFOURGE}|$  /(1 + ($z_\text{CANDELS}$+$z_\text{ZFOURGE})/2)$\,$>$\,=\,0.05, after the exclusion of AGS35, whose redshift in the ZFOURGE catalog ($z$\,=\,9.48) is much higher than that given in the CANDELS catalog, $z$\,=\,2.99.

Six galaxies in the Supplementary catalog have a spectroscopic redshift measurement: AGS26, $z_{sp}$\,=\,1.619 from VLT/FORS2 \citep{Vanzella2008}. AGS29, $z_{sp}$\,=\,1.117 from the VIMOS VLT Deep Survey \citep{Lefevre2013}. AGS30, $z_{sp}$\,=\,0.65 from the HST/ACS slitless grism spectroscopy of the PEARS program \citep{Ferreras2009}. AGS36, $z_{sp}$\,=\,0.646 from the Arizona CDFS Environment Survey (ACES), spectroscopic redshift survey of the Chandra Deep Field South (CDFS) using IMACS on the Magellan-Baade telescope \citep{Cooper2012} and confirmed by the VIMOS VLT Deep Survey \citep{Lefevre2013}. AGS37, $z_{sp}$\,=\,1.956 determined using the \textit{Spitzer} Infrared Spectrograph \citep{Wuyts2009, Fadda2010} and confirmed with the 3D-HST survey \citep{Momcheva2016}. AGS38, $z_{sp}$\,=\,1.314 determined with VLT/FORS2 \citep{Vanzella2008}.

We note that three additional spectroscopic redshifts have been reported for galaxies in the Main Catalog since the publication of \citetalias{Franco2018}. AGS6, previously reported at $z$\,=\,3.00, has been observed by the ALMA Spectroscopic Survey Large Program (ASPECS-LP; \citealt{Decarli2019}) in the \textit{Hubble} Ultra Deep Field, giving a $z_{sp}$\,=\,2.698. This spectroscopic redshift confirms the redshift also found by MUSE, at the same position \citep{Boogaard2019}. AGS18, previously reported at $z$\,=\,2.794, has also been observed in the ASPECS-LP survey, giving a $z_{sp}$\,=\,2.696. This spectroscopic redshift again confirms the one found by MUSE at the same position \citep{Boogaard2019}. AGS21, previously reported at $z$\,=\,3.76, has been observed by the VIMOS multi-object spectrograph of the VLT as part of the VANDELS survey \citep{Pentericci2018, McLure2018} and measured to have $z_{sp}$\,=\,3.689.

In the following, we will adopt for each source \textit{(i)} the spectroscopic redshift when available, otherwise \textit{(ii)} the photometric redshift from the ZFOURGE catalog (except for AGS35 for which we use the CANDELS redshift). These redshifts are given in Table~\ref{general_properties}. 

The stellar masses of the Main and Supplementary catalogs have been taken from the ZFOURGE catalog (except AGS35, for the reasons given above and for the large $z$ discrepancy). They were multiplied by a factor of 1.7 to scale them from the Chabrier IMF to a Salpeter IMF. Both catalogs provide globally consistent stellar masses with no systematic offset, the median of the ratio M$_{\star,\text{CANDELS}}$/M$_{\star, \text{ZFOURGE}}$ is 1.06 (see Fig.~\ref{compare_z}-right). 

For galaxies for which new redshift information has been obtained compared to the redshifts given in \cite{Straatman2016}, we derived new stellar masses to replace those given in the ZFOURGE catalog. For the sake of coherence and homogeneity, we used a similar technique to the one used for the ZFOURGE catalog. We used the code \texttt{FAST++}\footnote{Publicly available at \url{https://github.com/cschreib/fastpp}} based on \texttt{FAST} \citep{Kriek2009}. Stellar masses have been derived from models assuming exponentially declining star-formation histories and a dust attenuation law as described by \cite{Calzetti2000}.

\begin{table}\footnotesize     
\centering          
\begin{tabular}{ccc}     
\hline 
ID & $z$ & M$_\star$ [M$_\odot$]\\
\hline
\hline
AGS4  & 3.556$^{sp}$          & 11.09$_{-0.18}^{+0.06}$\\
AGS11 & 3.47                  & 10.24$_{-0.00}^{+0.75}$\\
AGS15 & 3.47                  & 10.56$_{-0.41}^{+0.01}$\\
AGS17 & 3.467$^{sp}$          & 10.52$_{-0.06}^{+0.40}$\\                
\hline                  
\end{tabular}
\caption{Refinement of redshift and stellar mass measurements for optically dark galaxies presented in the Main Catalog. IDs, redshifts and stellar masses for the optically dark galaxies that have been presented in \citetalias{Franco2018}, for which new measurements have now made it possible to refine the results (see \citealt{Zhou2020}). Spectroscopic redshifts have been flagged with an "sp" superscript.}\label{table_HST_dark}  
\end{table}

To ensure homogeneity of our results, we compared the stellar masses obtained using this technique with the ZFOURGE stellar masses (before modification of the redshift) for our sample of galaxies. We find a median difference of $\sim$8\%. 

\begin{figure*}
\includegraphics[width=\hsize]{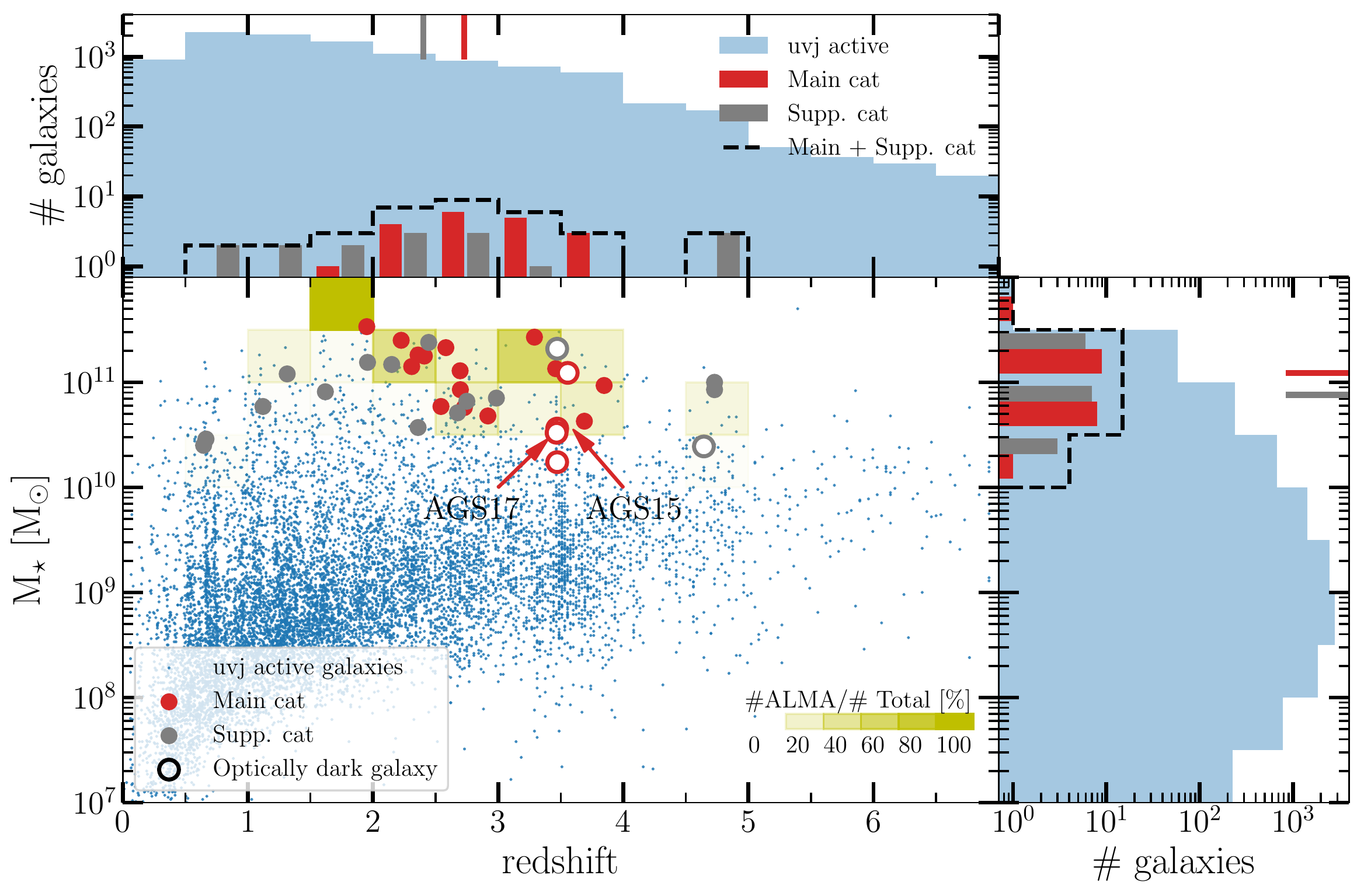}
      \caption{Stellar mass as a function of redshift for the galaxies detected in \citetalias{Franco2018} (red points) and in this work (gray points). For comparison, the distribution of all the galaxies, listed in the ZFOURGE catalog, in the same field of view is given in blue. Only UVJ active galaxies are shown. For each bin of redshift ($\Delta$$z$\,=\,0.5) and stellar mass (log$_{10}$(M$_{\star}$/M$_\odot$)\,=\,0.5), the fraction of sources detected by ALMA compared to the UVJ active galaxies in GOODS-ALMA is indicated with a yellow shading. The optically dark galaxies for which redshifts and masses have been derived are represented by open circles. The upper panel shows the compared distribution of redshift between all the UVJ active galaxies in GOODS-ALMA and the ALMA-detected galaxies while the right panel shows the stellar mass distribution. The median redshift and stellar mass are shown in these two panels. The median redshift is 2.40 for the galaxies presented in this paper, compared to 2.73 in \citetalias{Franco2018}, while the median stellar mass is 7.6$\times$10$^{10}$M$_\odot$ in this study, compared to 1.2$\times$10$^{11}$M$_\odot$ in \citetalias{Franco2018}.}
         \label{mass_redshift_combined}
   \end{figure*}

In addition, the masses and redshifts of the four optically dark galaxies presented in \citetalias{Franco2018} have recently been refined and will be presented in \cite{Zhou2020}, and are also summarized in Table~\ref{table_HST_dark}. The distributions of stellar mass and redshift of all the ALMA detections in GOODS-ALMA (Main Catalog and Supplementary Catalog) can be seen in Fig.~\ref{mass_redshift_combined} and are listed in Table~2 in \cite{Franco2020}.

\subsection{Flux and size measurements}\label{Flux_and_size_measurement}

Flux densities of the Supplementary Catalog sources were measured by fitting the light profiles with a circular Gaussian in the \textit{uv}-plane, using \texttt{uvmodelfit} in \texttt{CASA} \citep{McMullin2007}. Due to the relatively low S/N (3.5\,$<$\,S/N\,$<$\,4.8), we chose to fit a circular Gaussian rather than an asymmetric Gaussian, in order to limit the number of free parameters. We use the formula given by \cite{Marti-Vidal2012} (as in \citetalias{Franco2018}) to determine the minimum size that can be reliably measured in the \textit{uv}-plane by an interferometer, as a function of the S/N of the source:

\begin{equation}
 \theta_{min}\,\simeq\,0.88\frac{\theta_{beam}}{\sqrt{\text{S/N}}} \ .
 \label{Marti}
\end{equation}

\begin{figure}
   \centering
\includegraphics[width=1.\hsize]{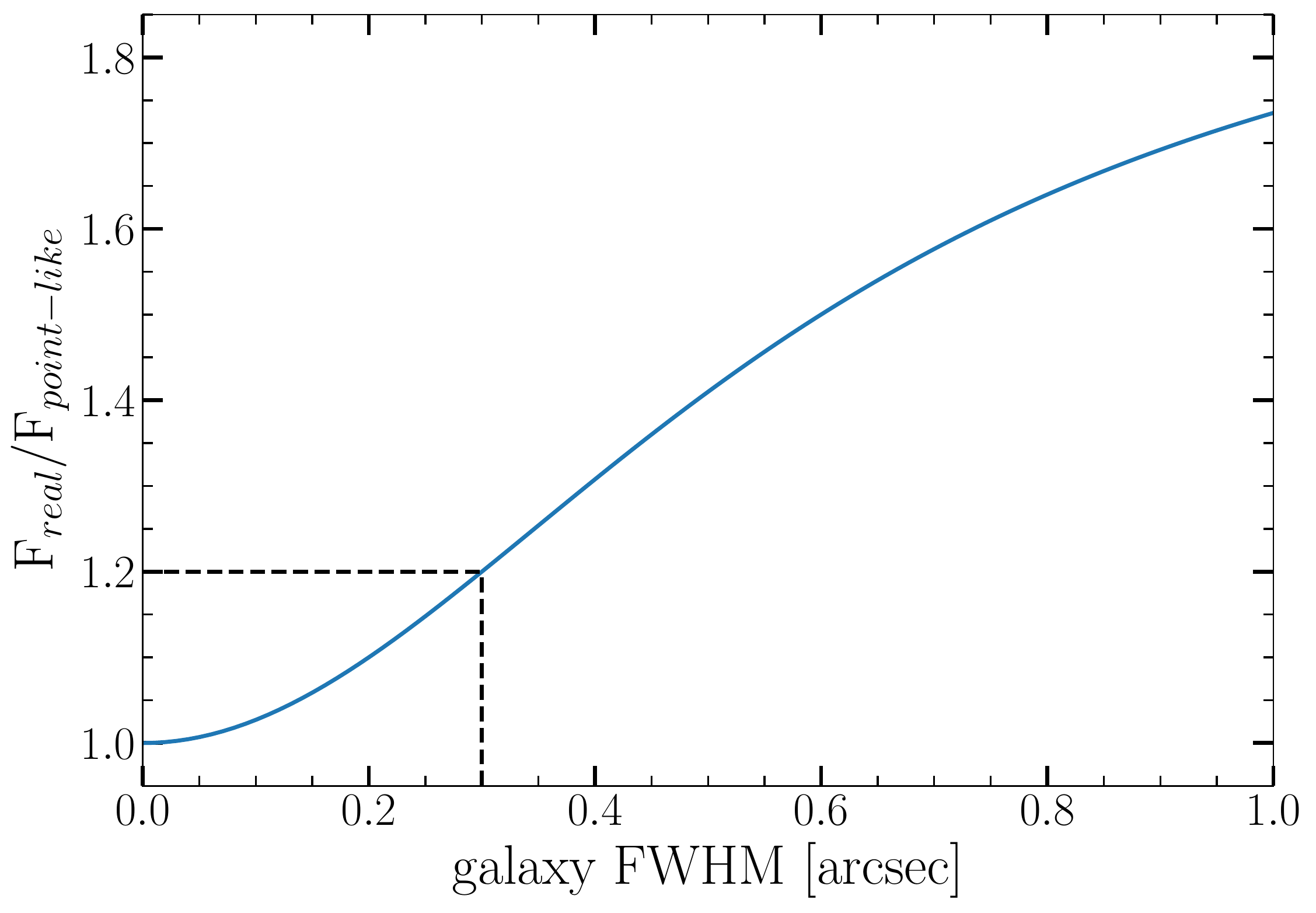}
\caption{Underestimation of the flux when assuming a point-like source instead of the real size of the galaxy. For example, a source with an intrinsic FWHM of 0\farcs3 will be underestimated by 20\% (dashed-line).}
         \label{convolution_beam}
   \end{figure}

\begin{figure}
   \centering
   \includegraphics[width=\hsize]{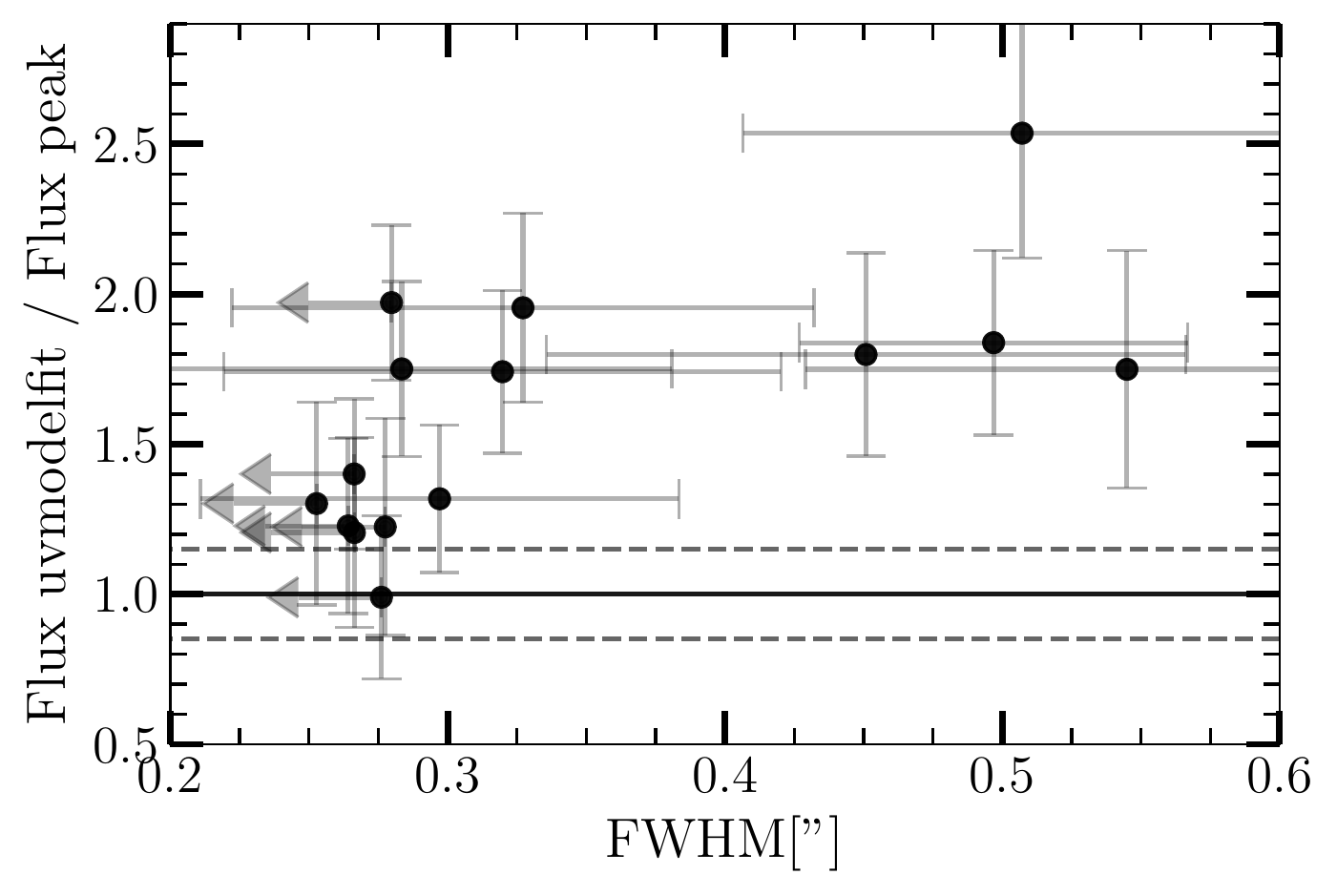}
      \caption{Ratio between the flux extracted using \texttt{uvmodelfit} in \texttt{CASA} and the peak flux as a function of the size of the galaxy for the supplementary catalog. When the measured size is below the reliable size measurement limit (see Eq.~\ref{Marti}), we consider the size given by Eq.~\ref{Marti} as an upper limit (see Table~\ref{Table_theta}). The horizontal solid line indicates flux \texttt{uvmodelfit}\,=\,peak flux. The dotted lines show a 15 percent deviation from this equality. For galaxies larger than 0\farcs25, the approximation of a point source is no longer valid, and we assume the flux value derived from \texttt{uvmodelfit} for these galaxies.}
         \label{flux_size}
   \end{figure}

To calculate $\theta_{min}$, we use the S/N of the sources in the tapered image at 0\farcs60, and therefore $\theta_{beam}$\,=\,0\farcs60. For galaxies where the circular Gaussian fit in the \textit{uv}-plane gave a size (FWHM) smaller than $\theta_{min}$ (the size limit given by Eq.~\ref{Marti}), we take the galaxy to be unresolved, and therefore set the size upper limit of this galaxy to $\theta_{beam}$ (see Table~\ref{Table_theta}), and use the peak flux density measured on the direct image. Assuming these sources are point-like is expected to lead to slightly underestimated flux densities, as the typical size measured for distant ALMA galaxies is on average close to 0$\farcs$3 (\citealt{Simpson2015b, Ikarashi2017, Elbaz2017}). Assuming a point-like source for a real size extension of 0$\farcs$3 FWHM would lead to an underestimation of the real flux density by a factor F$_{\nu,\rm real}$/F$_{\nu,\rm peak}$\,=\,1.2 (see Fig.~\ref{convolution_beam}). In the absence of a robust size measurement, we decided to keep the peak flux values keeping in mind that they may be lower by about 20\,\%. Using the measurements coming out of \texttt{uvmodelfit} would lead to larger uncertainties for those sources with no reliable size measurement. For galaxies whose sizes measured using \texttt{uvmodelfit} are larger than the size limit given by Eq.~\ref{Marti}, the peak flux approximation is no longer valid and we use the integrated flux given by \texttt{uvmodelfit} (see Fig.~\ref{flux_size}).

\begin{table}\footnotesize     
\centering          
\begin{tabular}{c c c c }     
\hline 
ID & $\theta_\texttt{uvmodelfit}$& $\theta_\text{lim}$& $\theta_\text{final}$\\
\hline
\hline
 AGS24 & 0.06 & 0.27 & $<$0.27 \\
 AGS25 & 0.12 & 0.25 & $<$0.25 \\
 AGS26 & 0.30 & 0.25 & 0.30 \\
 AGS27 & 0.54 & 0.27 & 0.54 \\
 AGS28 & 0.50 & 0.26 & 0.50 \\
 AGS29 & ...  & 0.28 & $<$0.28 \\
 AGS30 & ...  & 0.26 & $<$0.26 \\
 AGS31 & ...  & 0.27 & $<$0.27 \\
 AGS32 & 0.33 & 0.27 & 0.33 \\
 AGS33 & 0.51 & 0.27 & 0.51 \\
 AGS34 & ...  & 0.27 & $<$0.27 \\
 AGS35 & 0.45 & 0.27 & 0.45 \\
 AGS36 & 0.23 & 0.28 & $<$0.28 \\
 AGS37 & 0.28 & 0.28 & 0.28 \\
 AGS38 & 0.32 & 0.28 & 0.32 \\
 AGS39 & 0.25 & 0.28 & $<$0.28 \\
\hline                  
\end{tabular}
\caption{Table of sizes (FWHM) measured with \texttt{uvmodelfit} and reliable size measurement limits given by \cite{Marti-Vidal2012}. The last column gives the adopted size. If $\theta_\texttt{uvmodelfit}$ $>$ $\theta_\text{lim}$, we take $\theta_\texttt{uvmodelfit}$ as the final size. If $\theta_\texttt{uvmodelfit}$ $<$ $\theta_\text{lim}$, we use the upper limit $\theta_\text{lim}$. The absence of size indicates a non-convergence of \texttt{uvmodelfit}.}
\label{Table_theta}  
\end{table}

In addition, as we are exploring a low S/N regime, we have verified the  impact of flux boosting on our results. Using the same set of injected sources as those we used to measure the positional differences of the ALMA sources (see Section \ref{subSect::ALMA_Supplementary_Catalog}), we compared the recovered flux to the injected flux. At a recovered flux of 0.9\,mJy which corresponds to the characteristic flux of our sample, the flux boosting reaches 10\%. However, as the dispersion of the flux boosting is higher than the correction to be applied and because the correction to be applied is low, we decided not to apply the flux boosting correction to our results.

\section{Comparison of the properties of the ALMA galaxies from the Main and Supplementary Catalogs}\label{Sect::Properties}

\subsection{Redshifts}

The redshifts of the Supplementary Catalog cover a wider range ($z$\,=\,0.65 -- 4.73) than the sources of the Main Catalog ($z$\,=\,1.95 -- 3.85). While there are no galaxies with a redshift greater than $z$\,=\,4 in the Main Catalog, these galaxies make up 19\,\% of the Supplementary Catalog (3/16, see Fig.~\ref{mass_redshift_combined}). At the other extreme, none of the Main Catalog sources were below $z$\,=\,1.9 whereas 38\,\% (6/16) of the sources in the Supplementary Catalog are found in this lower redshift range. These low-redshift galaxies reduce the median redshift $z_{med,SC}$\,=\,2.40 compared to that of the Main Catalog $z_{med,MC}$\,=\,2.73. 
This median redshift is also similar to that of \cite{Stach2019}, who derive a median redshift of 2.61\,$\pm$\,0.09 in an ALMA follow-up of the SCUBA-2 Cosmology Legacy Survey in the UKIDSS/UDS field. We found no significant correlation between redshift and flux density (Spearman's correlation coefficient $\rho$\,=\,0.30).

\subsection{Stellar masses}

All galaxies detected in GOODS-ALMA have stellar masses greater than M$_{\star}$\,=\,2$\times$10$^{10}$M$_\odot$. The median stellar mass of galaxies from the Supplementary Catalog, M$_{\star}^{\rm Supp}$\,=\,7.6$\times$10$^{10}$M$_\odot$, is 1.6 times smaller than that of galaxies in the Main Catalog, M$_{\star}^{\rm Main}$\,=\,1.2 $\times$10$^{11}$M$_\odot$. Hence by pushing down the ALMA detection limit using the IRAC catalog, we have reached more normal galaxies, with less extreme stellar masses.

We can now compare the galaxies detected by GOODS-ALMA, combining the Main and Supplementary Catalogs, to their parent sample of distant star-forming galaxies taken from the ZFOURGE catalog after selecting only the UVJ active (star-forming) galaxies (\citealt{Williams2009}), using the same definition as in \citetalias{Franco2018} (see Fig.~\ref{mass_redshift_combined}). GOODS-ALMA detects nearly half (46\%, 6/13) of the most massive star-forming galaxies with log$_{10} $(M$_{\star}$$/$M$_\odot$)\,=\,11--12 in the range 2\,$<$\,$z$\,$<$\,2.5. Pushing further in redshift to 2.5\,$<$\,$z$\,$<$\,3, GOODS-ALMA also detects nearly half of the star-forming galaxies with log$_{10}$ (M$_{\star}$$/$M$_\odot$)\,=\,10.7--11 (44\%, 7/16). At even higher redshifts, 3\,$<$\,$z$\,$<$\,4, GOODS-ALMA detects 38\% (3/8) of the most massive galaxies (log$_{10}$ (M$_{\star}$$/$M$_\odot$)\,=\,11--12). In total, GOODS-ALMA detects approximately 38\% (12/32) of the most massive star-forming galaxies with redshifts 2\,$<$\,$z$\,$<$\,4 (log$_{10}$ (M$_{\star}$$/$M$_\odot$)\,=\,11 -- 12), this by taking into account the two optically dark present in this interval. 

\subsection{Sizes}
   \begin{figure}
   \centering
   \includegraphics[width=\hsize]{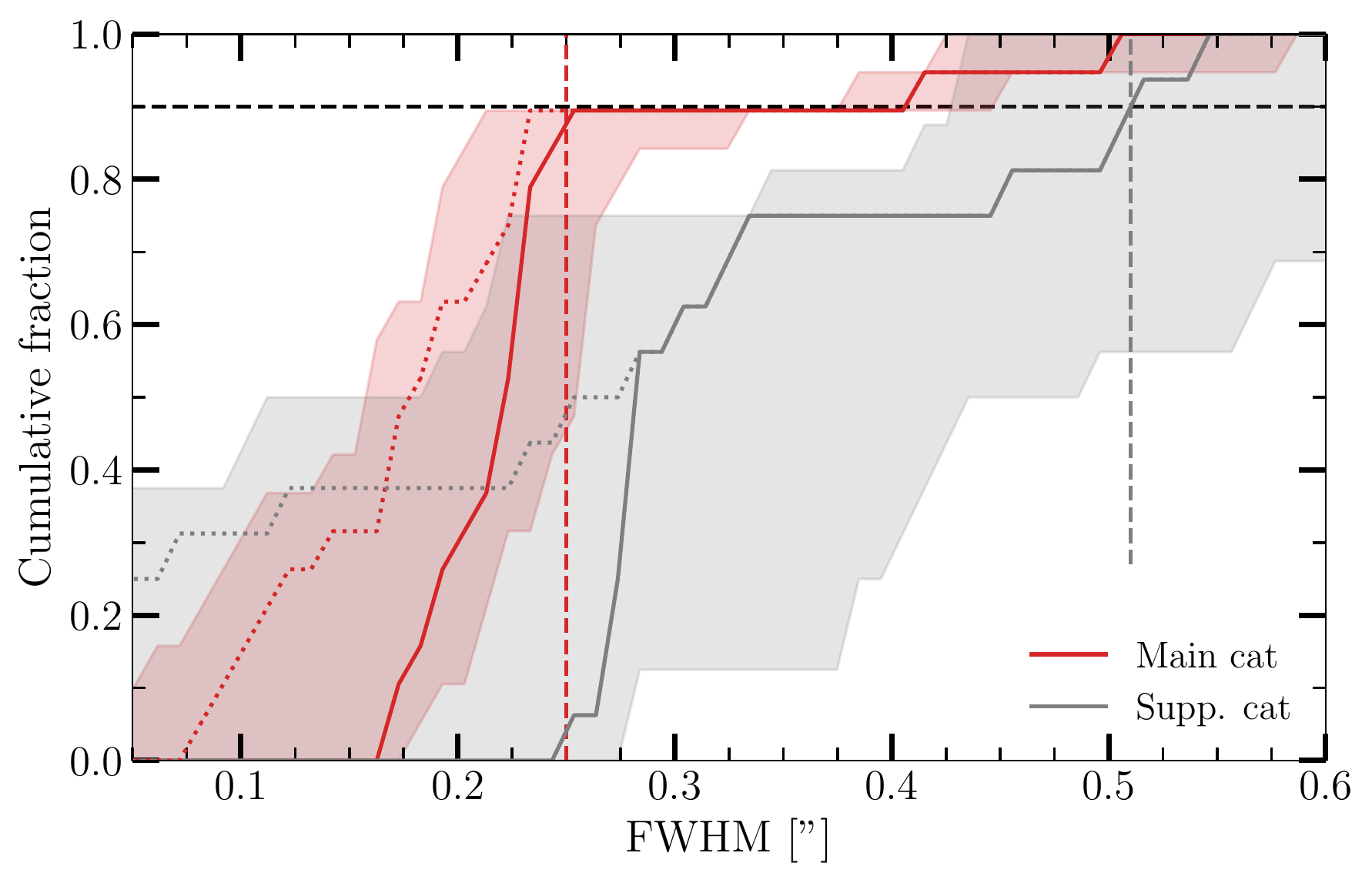}
      \caption{Cumulative fraction of sources with a FWHM below a given size for the main (red) and the supplementary catalog (gray). These sizes are computed by fitting the ALMA detections with a circular Gaussian in the \textit{uv}-plane using \texttt{uvmodelfit} in \texttt{CASA}. The dotted lines refer to the sizes given by \texttt{uvmodelfit}, while the solid lines take into account  take into account the upper limits given by Eq.~\ref{Marti}. The shaded areas correspond to the integration of the individual uncertainties of the sizes of each detection.}
         \label{size_plan_uv}
   \end{figure}

The sizes of the sources of the Main and Supplementary catalogs were derived by fitting a circular Gaussian in the \textit{uv}-plane using \texttt{uvmodelfit} in \texttt{CASA}. We find that by pushing down the detection limit to 3.5-$\sigma$ using IRAC and VLA, we have been able to identify galaxies with nearly twice larger ALMA sizes than those in the Main catalog. The median ALMA 1.1\,mm FWHM of the galaxies in the Supplementary Catalog is indeed 0$\farcs$32 as compared to 0$\farcs$18 for the galaxies in the Main Catalog. When accounting for the redshift of the sources, we find that the physical circularized half-light radius of the new sources in the present Supplementary Catalog (R$_{1/2}$\,=\,FWHM/2) is 1.3 kpc as compared to only 0.65 kpc for the Main Catalog. If we take into account the fact that the Supplementary sources exhibit stellar masses that are half of those of source in the Main catalog, this implies that by pushing down the ALMA detection limit using IRAC and VLA catalogs, we were able to identify lower stellar mass galaxies in which dust-enshrouded star formation extends over twice larger sizes. 

In Fig.~\ref{size_plan_uv}, we show the cumulative fraction of sources with a major axis below a given size for the Main (red) and Supplementary (gray) catalogs. This figure clearly shows that the galaxies detected in the Main catalog are generally more compact than those in the Supplementary catalog: 90\% of galaxies in the Main Catalog have a FWHM below 0$\farcs$25, whereas 90\,\% of the galaxies found in the Supplementary Catalog have a FWHM of less than 0$\farcs$50. Moreover, 40\% of sources in the Supplementary Catalog have a FWHM size above 0$\farcs$30 arcsec.

This shows that while the projected sizes of dust-enshrouded star formation probed by ALMA are globally small for massive and distant galaxies, the new sources that we present here in the Supplementary Catalog do not extend the sample to much lower flux densities but to sources with a wider extension of the dust emission. This explains in part why these sources were not detected in the Main Catalog. Although their integrated flux densities may be equal (and sometimes higher) than sources in the Main Catalog, this flux is diluted into several beams and therefore drops below the detection limit for the central beam. We recall that this increase in the ALMA sizes measured in the Supplementary Catalog remains such that globally the ALMA emission extends over much smaller sizes than their $H$-band sizes, confirming that the ALMA sources are particularly compact at 1.1\,mm \cite[e.g.,][]{Chen2015, Simpson2015, Rujopakarn2016, Elbaz2017, Calistro_Rivera2018, Franco2020}.

\subsection{How complete is the Main plus Supplementary Catalog?}

The average noise in the GOODS-ALMA image is RMS\,=\,0.182\,mJy, hence the 3.5-$\sigma$ limit of the Supplementary Catalog converts into a detection limit of about 0.64\,mJy. We note that since the RMS of the noise varies across the image because it is subdivided in 6 slices taken at different epochs, sources may be detected below 0.64\,mJy (e.g., a source was detected at 0.55\,mJy).

The various studies that have carried out millimetric source counts \cite[e.g.,][]{Hatsukade2013,Oteo2015,Aravena2016,Umehata2017,Fujimoto2017,Dunlop2017,Franco2018,Hatsukade2018} allow us to estimate an expected galaxy surface density that varies between $\sim$2000 and $\sim$3500 galaxies/deg$^{2}$ above 0.65\,mJy at 1.1\,mm. Over the size of 69.5\,arcmin$^2$ of the GOODS-ALMA survey, this amounts to an estimated number of sources ranging between 39 and 48. By comparison, we have now extended the number of detections in GOODS-ALMA to 35 galaxies. This number is not far from the expected value, especially when one accounts for cosmic variance, and suggests that the present sample may be more than 70\,\% complete above 0.65\,mJy. In addition, the advantage of working on a contiguous and homogeneous mosaic of $\sim$\,70 arcmin$^2$ is to be able to push beyond the detection limit of individual galaxies by performing a stacking analysis at the position of galaxies detected at other wavelengths. This will allow us to quantify the density of star formation at high redshift and perform statistical tests. These stacking studies are in progress and will lead to several papers in preparation.

\section{Conclusions}

Using IRAC and VLA (combined with deep $Ks$ images), we are able to explore the presence of galaxies detected at 1.1\,mm with ALMA down to the 3.5-$\sigma$ limit. This was done despite the extremely large number of independent beams in the ALMA image, even after tapering from 0$\farcs$29 to 0$\farcs$6.

In order to avoid introducing spurious associations, we restricted the new sample to ALMA detections with either an IRAC counterpart closer than 0$\farcs$3 or closer than 0$\farcs$7 but with a radio counterpart as well. In two cases, we used the $K$-band image to deconvolve IRAC sources that were missed by previous studies because of their close proximity to bright IRAC neighbors. These two galaxies do not exhibit any counterpart in the HST images, hence they are optically dark, but both present a radio counterpart. In order to minimize the chance of false associations, we have deeply investigated the astrometry between the different instruments used in this paper. We used a comparison of nearly 400 galaxies in common between HST and Pan-STARRS in the GOODS-ALMA field. We show that the astrometry of the HST image does not only suffer from a global astrometric shift, as already discussed in previous papers, but also a local shift that results in the equivalent of a distortion map that was artificially introduced in the process of building the mosaic of the GOODS--South HST image. We present a solution to correct for this distortion and use this correction in our identification of counterparts. We note that in some cases, the absence of this correction led previous studies to attribute the wrong counterpart to ALMA detections.

In total we find 16 galaxies in the Supplementary Catalog that bring the total sample of GOODS-ALMA 1.1\,mm sources to 35 galaxies. This number is between 70 and 90\,\% of the predicted number of galaxies expected to be detected at 1.1\,mm above 0.65\,mJy as derived from existing millimeter number counts. We now detect in GOODS-ALMA between a third and half of the most massive star-forming galaxies (log$_{10}$(M$_{\star}$$/$M$_\odot$)\,=\,11 -- 12), depending on the redshift range within 2\,$<$\,$z$\,$<$\,4. 

The redshift range of the Supplementary Catalog covers a wider range ($z$\,=\,0.65 -- 4.73) than the sources of the Main Catalog ($z$\,=\,1.95 -- 3.85). The median redshift of the Supplementary Catalog $z_{med,SC}$\,=\,2.40 is slightly lower than that of the Main Catalog $z_{med,MC}$\,=\,2.73 due to the presence of low-redshift galaxies. The typical physical size of the new sources in the present Supplementary Catalog (1.3 kpc) is twice larger than that of the Main Catalog sources (0.65 kpc). The lower surface brightness of these sources explains partly why they were not detected in the Main Catalog. Hence, pushing down the ALMA detection limit using IRAC and VLA allowed us to reach galaxies with lower stellar masses than in the Main Catalog (median stellar mass M$_{\star}$\,=\,7.6$\times$10$^{10}$M$_\odot$) in which dust-enshrouded star formation extends over twice larger sizes. However, this increase in the ALMA sizes is not large enough to question the fact that the ALMA emission globally extends over much smaller sizes than the $H$-band light, confirming that the ALMA sources are particularly compact at 1.1\,mm. The properties of the galaxies presented in this paper are discussed in more detail in \cite{Franco2020}.

\section{Acknowledgements}
We thank the anonymous referee for the insightful comments and suggestions that improved the clarity and quality of this work. M.F. acknowledges support from the UK Science and Technology Facilities Council (STFC) (grant number ST/R000905/1). B.M. acknowledges support from the Collaborative Research Centre 956, sub-project A1, funded by the Deutsche Forschungsgemeinschaft (DFG) -- project ID 184018867. L.Z. acknowledges the support from the National Key R\&D Program of China (No. 2017YFA0402704, No. 2018YFA0404502),  the National Natural Science Foundation of China (NSFC grants 11825302, 11733002 and 11773013) and China Scholarship Council (CSC). R.D. gratefully acknowledges support from the Chilean Centro de Excelencia en Astrof\'isica y Tecnolog\'ias Afines (CATA) BASAL grant AFB-17000. GEM acknowledges support from  the Villum Fonden research grant 13160 “Gas to stars, stars to dust: tracing star formation across cosmic time”, the Cosmic Dawn Center of Excellence funded by the Danish National Research Foundation and  the ERC Consolidator Grant funding scheme (project ConTExt, grant number No. 648179). MP is supported by the ERC-StG 'ClustersXCosmo', grant agreement 71676. DMA acknowledges support from the Science and Technology Facilities Council (ST/P000541/1; ST/T000244/1). This work was supported by the Programme National Cosmology et Galaxies (PNCG) of CNRS/INSU with INP and IN2P3, co-funded by CEA and CNES. This paper makes use of the following ALMA data: ADS/JAO.ALMA\#2015.1.00543.S. ALMA is a partnership of ESO (representing its member states), NSF (USA) and NINS (Japan), together with NRC (Canada), MOST and ASIAA (Taiwan), and KASI (Republic of Korea), in cooperation with the Republic of Chile. The Joint ALMA Observatory is operated by ESO, AUI/NRAO and NAOJ. 

\bibliographystyle{aa}
\bibliography{biblio}

\onecolumn
\begin{appendix}
\section{Multiwavelength postage-stamps}

 \begin{figure*}[h!]
 \centering
 \vspace{-0.6cm}
\includegraphics[width=.79\hsize]{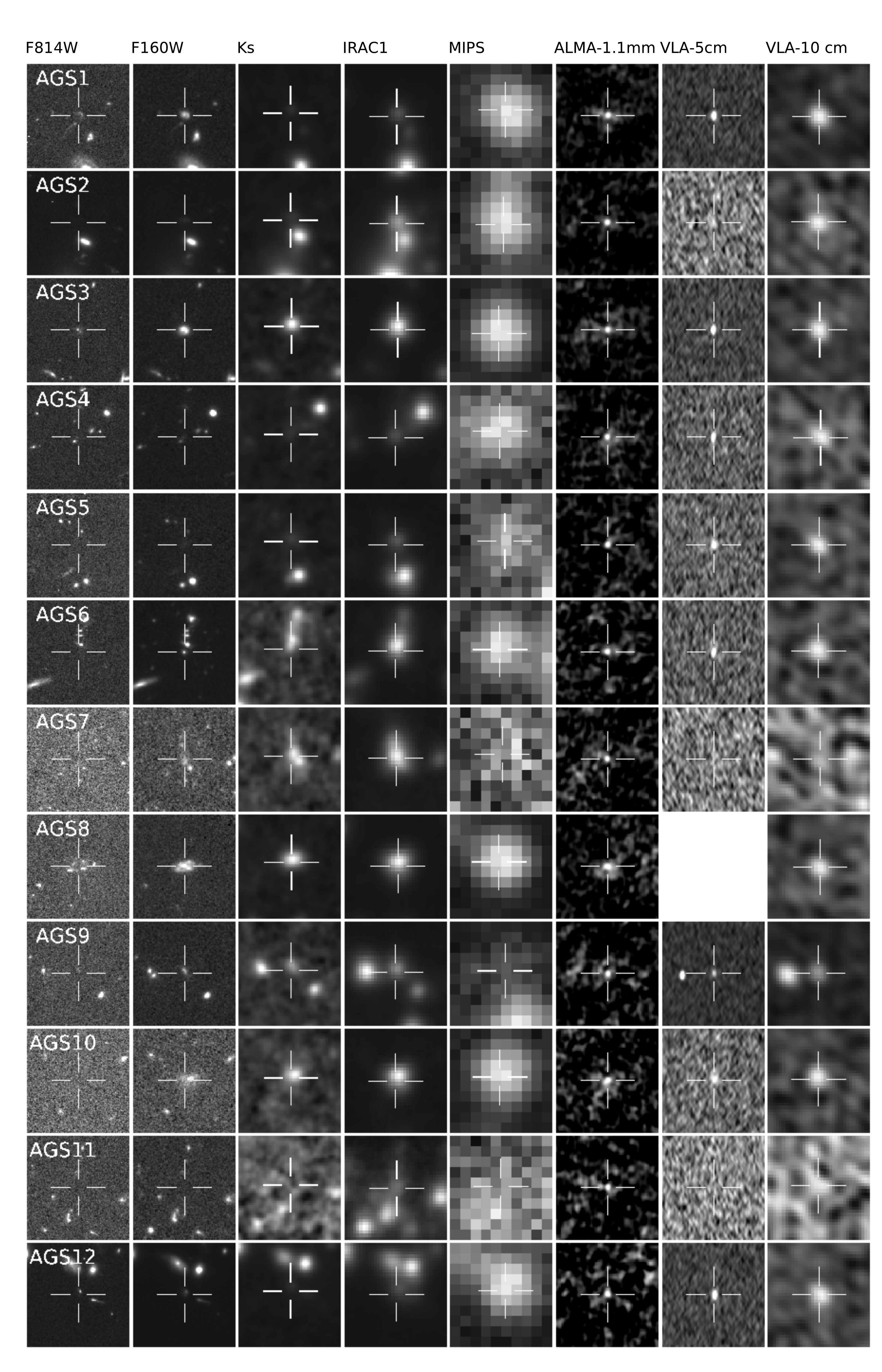}
\caption{Postage-stamps (10\,$\times$\,10 arcseconds), centered on the position of the ALMA detection at different wavelengths. From left to right : HST-WFC3 (F814W, F160W), ZFOURGE (K$_s$), \textit{Spitzer}-IRAC channel 1 (3.6 $\mu$m), \textit{Spitzer}-MIPS (24$\mu$m), ALMA band 6 (1.1\,mm), VLA (5 and 10\,cm). Blank images mean that the source is out of the field of view of the instrument. The white cross indicates the position of the ALMA detection. The north is up and the east is left.}
\label{multiwav_main_cat}
\end{figure*}

  \begin{figure*}
  \centering
  \includegraphics[width=0.79\hsize]{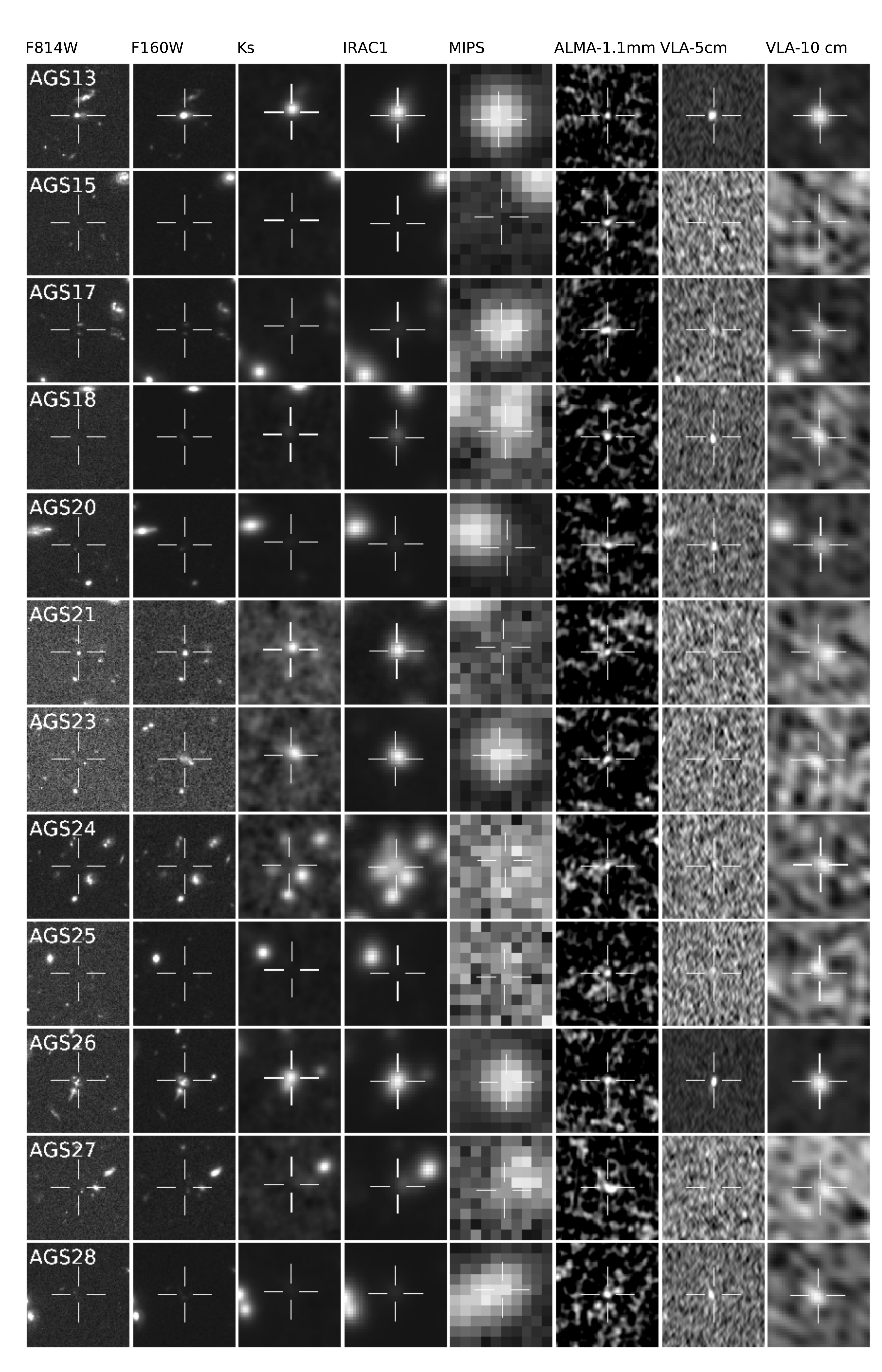}
      \caption{(continued)}
  \end{figure*}
   
    \begin{figure*}
  \centering
  \includegraphics[width=0.79\hsize]{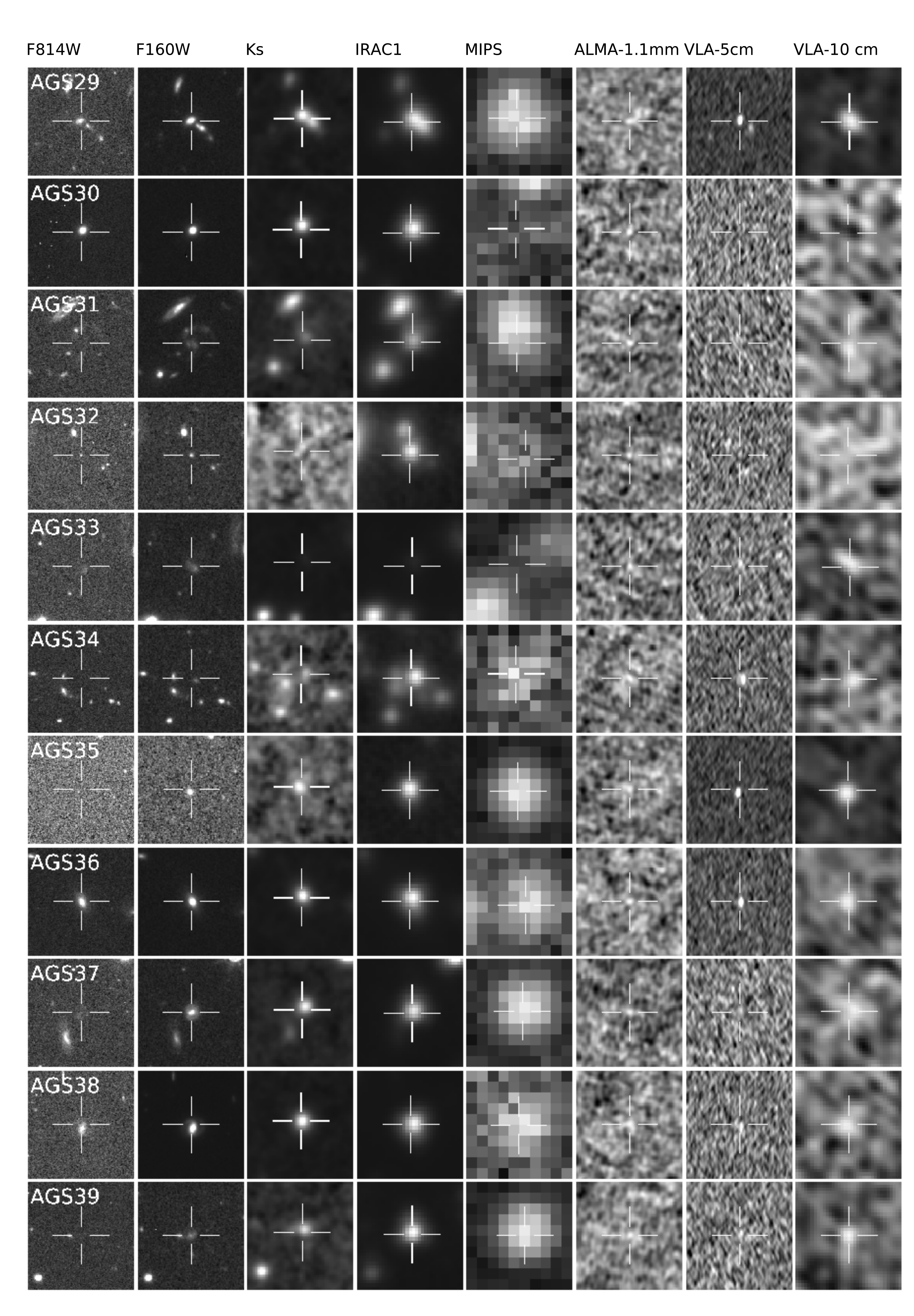}
      \caption{(continued)}
  \end{figure*}
    \end{appendix}
\end{document}